\documentclass[a4paper,pra,twocolumn,showpacs,superscriptaddress,groupedaddress]{revtex4}
\usepackage{ae}
\usepackage{physics}
\usepackage[T1]{fontenc}
\usepackage[ansinew]{inputenc}
\usepackage{amsmath}
\usepackage{amssymb}
\usepackage[caption=false]{subfig}
\usepackage{multirow}
\usepackage{array}
\usepackage[]{graphicx}
\usepackage{wrapfig}
\usepackage{makecell}

\usepackage{color}
\usepackage[colorlinks]{hyperref}
\usepackage{lscape}
\graphicspath{ {./images1/} }

\begin{document}
\title{Nonlocality of tripartite orthogonal product states}
\author{Atanu Bhunia}
\email{atanu.bhunia31@gmail.com}
\affiliation{Department of Applied Mathematics, University of Calcutta, 92, A.P.C. Road, Kolkata- 700009, India}
\author{Indrani Chattopadhyay}
\email{icappmath@caluniv.ac.in}
\affiliation{Department of Applied Mathematics, University of Calcutta, 92, A.P.C. Road, Kolkata- 700009, India}
\author{Debasis Sarkar}
\email{dsarkar1x@gmail.com, dsappmath@caluniv.ac.in}
\affiliation{Department of Applied Mathematics, University of Calcutta, 92, A.P.C. Road, Kolkata- 700009, India}

\begin{abstract}
 Local distinguishability of orthogonal product states is an area of active research in quantum information theory. However, most of the relevant results about local distinguishability found in bipartite quantum systems and very few are known in multipartite systems. In this work, we construct a locally indistinguishable subset in  ${\mathbb{C}}^{2d}\bigotimes{\mathbb{C}}^{2d}\bigotimes{\mathbb{C}}^{2d}$, $d\geq2$  that contains  $18(d-1)$ orthogonal product states. Further, we generalize our method to arbitrary tripartite quantum systems ${\mathbb{C}}^{k}\bigotimes{\mathbb{C}}^{l}\bigotimes{\mathbb{C}}^{m}$. This result enables us to understand further the role of nonlocality without entanglement in multipartite quantum systems. Finally, we prove that a three-qubit GHZ state is sufficient as a resource to distinguish each of the above classes of states.
\end{abstract}
\date{\today}
\maketitle
\section{INTRODUCTION}
Many global operations cannot be accomplished using only local operations and classical communications (LOCC). Therefore, researchers are interested to investigate the restrictions of quantum operators that can be implemented by LOCC. In state level, the local distinguishability of quantum states plays an important role in studying the restrictions of LOCC \cite{1}. Whereas, the local indistinguishability of pure orthogonal product states exhibits the phenomenon of nonlocality without entanglement \cite{1,2,3,4,5,6,7,8,9,10,11,12,13,14,15,16,17,18,19,20,21,22,23,24,25,26,27,28,29,30,31,32,33,34,35}. The nonlocality without entanglement(NLWE) can be viewed as a new striking example of the nonequivalence between the concept of quantum entanglement and that of quantum nonlocality. It was known that entanglement does not necessarily imply nonlocality in the sense of producing data that are incompatible with local realism. However, this new type of nonlocality implies that the converse does not hold either. Note that nonlocality is not understood here as the incompatibility with local realism, but instead as the advantage of a joint measurement with respect to LOCC.\\  
When restricted to LOCC, two distant parties Alice and Bob generally cannot distinguish among bipartite states as they can with global operations. This is also true when some set of states are pairwise orthogonal (as pairwise orthogonal states can be perfectly distinguished by global operations). Projective measurements on composite systems are in general global operations that cannot be implemented by LOCC. Hence generally, pairwise orthogonal states cannot be perfectly distinguished by LOCC.\\
Walgate et al. \cite{36} showed that by using LOCC only any two pure orthogonal multipartite  states can be perfectly distinguished. The local indistinguishability of pairwise orthogonal set of multipartite states is a signature of nonlocality showed by those states\cite{36,37,38,39,40,41,42,43,44}. Since entanglement is deeply attached to nonlocality, one can assume that pairwise orthogonal product states would be perfectly distinguished by LOCC. However, this assumption is wrong. Bennett et al. \cite{1} showed that there exist a complete set of nine pure orthogonal product states in ${\mathbb{C}}^{3}\bigotimes{\mathbb{C}}^{3}$ which is indistinguishable by LOCC. So, the local distinguishability problem is not so simple when the number of states is more than two \cite{4,36}.\\
Quantum nonlocality without entanglement in a multipartite quantum system is still incompletely studied except for some special completely orthogonal product bases(COPB) \cite{1,2,4,5,36} and some unextendible  product bases(UPBs) \cite{3}. On the other hand, Bennett et al. exhibited a complete orthogonal product basis with eight members which cannot be distinguished by LOCC in ${\mathbb{C}}^{2}\bigotimes{\mathbb{C}}^{2}\bigotimes{\mathbb{C}}^{2}$. Niset and Cerf \cite{5} constructed a complete product basis for $n$ parties, each holding a system of at least $(n-1)$ dimensions which exhibits nonlocality without entanglement. On the other hand, Bennett et al. proved that a UPB cannot be perfectly distinguished by LOCC and gave an example of a UPB of three parties, which is known as SHIFTS UPB. DiVincenzo et al. \cite{3}, gave a generic method to construct a multipartite UPB with an even number of members.\\
In this paper, we construct a set of product states in ${\mathbb{C}}^{2d}\bigotimes{\mathbb{C}}^{2d}\bigotimes{\mathbb{C}}^{2d}$ which is LOCC indistinguishable. If a set of quantum states is LOCC distinguishable, then there exist a nontrivial measurement that can preserve the orthogonality of these states. If not, these quantum states must be LOCC indistinguishable. Here we construct a subset in  ${\mathbb{C}}^{2d}\bigotimes{\mathbb{C}}^{2d}\bigotimes{\mathbb{C}}^{2d}$, $d\geq2$  that contains $18(d-1)$ orthogonal product states that are locally indistinguishable. Next, we generalize our method to arbitrary tripartite quantum system ${\mathbb{C}}^{k}\bigotimes{\mathbb{C}}^{l}\bigotimes{\mathbb{C}}^{m}$. The results help us to understand the role of nonlocality without entanglement further in multipartite systems. Finally, it can be shown that a three-qubit GHZ state is sufficient as a resource to distinguish each of the above classes of states. 
 
\section{SOME LOCC INDISTINGUISHIBLE CLASSES IN TRIPARTITE SYSTEM}
In this section, we will construct a set of orthogonal product states in ${\mathbb{C}}^{2d}\bigotimes{\mathbb{C}}^{2d}\bigotimes{\mathbb{C}}^{2d}$ shared between three parties Alice, Bob and Charlie, which is LOCC indistinguishable.  For better understanding, we first give an example in  ${\mathbb{C}}^{6}\bigotimes{\mathbb{C}}^{6}\bigotimes{\mathbb{C}}^{6}$ and generalize the result for higher dimension. Next, we will provide various classes of states in tripartite quantum system which are LOCC indistinguishable. Here, we represent a quantum state $\ket{i+\overline{i+1}
}\ket{j}\ket{k}$ by a cuboid, where $\ket{i\pm\overline{i+1}
}=\frac{1}{\sqrt{2}}(\ket{i}\pm\ket{i+1}),$ for integer $i$.

Example 1: In \ \        ${\mathbb{C}}^{6}\bigotimes{\mathbb{C}}^{6}\bigotimes{\mathbb{C}}^{6}$ the set of 36 orthogonal product states\\
$$\begin{Bmatrix}
\begin{matrix}
\ket{2+3}\ket{0}\ket{3}\\
\ket{3}\ket{2+3}\ket{0}\\
\ket{0}\ket{3}\ket{2+3}
\end{matrix} & 
\begin{matrix}
\ket{2+3}\ket{2}\ket{5}\\
\ket{5}\ket{2+3}\ket{2}\\
\ket{2}\ket{5}\ket{2+3}
\end{matrix} & 
\begin{matrix} 
\ket{4\pm5}\ket{2}\ket{5}\\
\ket{5}\ket{4\pm5}\ket{2}\\
\ket{2}\ket{5}\ket{4\pm5}
\end{matrix}\\\\
\begin{matrix}
\ket{2\pm3}\ket{2}\ket{3}\\
\ket{3}\ket{2\pm3}\ket{2}\\
\ket{2}\ket{3}\ket{2\pm3}
\end{matrix} & 
\begin{matrix}
\ket{3\pm4}\ket{2}\ket{4}\\
\ket{4}\ket{3\pm4}\ket{2}\\
\ket{2}\ket{4}\ket{3\pm4}
\end{matrix}\\\\ 
\begin{matrix} 
\ket{1\pm2}\ket{1}\ket{3}\\
\ket{3}\ket{1\pm2}\ket{1}\\
\ket{1}\ket{3}\ket{1\pm2}
\end{matrix} & 
\begin{matrix}
\ket{0\pm1}\ket{0}\ket{3}\\
\ket{3}\ket{0\pm1}\ket{0}\\
\ket{0}\ket{3}\ket{0\pm1}
\end{matrix}
\end{Bmatrix}$$\\
depicted in FIG. 1, is LOCC indistinguishable.\\\\
\textit{Proof}: To distinguish those product states, some party has to start with a nontrivial orthogonality preserving measurement, say Alice (the first party). Since, the states are symmetrical with respect to parties, it is clear that if the states cannot be distinguished with Alice going first, then these states cannot be distinguished whether Bob or Charlie going first. Thus we only need to prove that these orthogonal product states cannot be distinguished by LOCC if Alice going first with orthogonality preserving positive-operator-valued  measurement(OPPOVM)\cite{4,8,10}.\\
Without loss of generality, suppose Alice goes first with a set of $6\times6$  positive-operator-valued measurement(POVM) elements $\{{{M_{Ai}}^{\dagger}M_{Ai}}\}$. 
Let us consider an $M_{A}$ be any matrix that preserves orthogonality of the above product states, where $M_A$ has the following representation under the basis elements $\{\ket{0},\ket{1},.....,\ket{5}\},$\\
$${M_{A}}^{\dagger}M_{A}=\begin{pmatrix}
m_{00} & m_{01} & . & . & . & m_{05}\\
m_{10} & m_{11} & . & . & . & m_{15}\\
. & . & . &   &  & .\\
. & . &   & . &  & .\\
. & . &   &   & . & .\\
m_{50} & m_{51} & .&.&.& m_{55}
\end{pmatrix}$$\\
Next, we prove that ${M_A}^{\dagger}M_A\propto{\alpha I},$  where ${I}$ is the identity operator.\\
\begin{figure}[h!]
	\centering
	\includegraphics[width=0.50\textwidth]{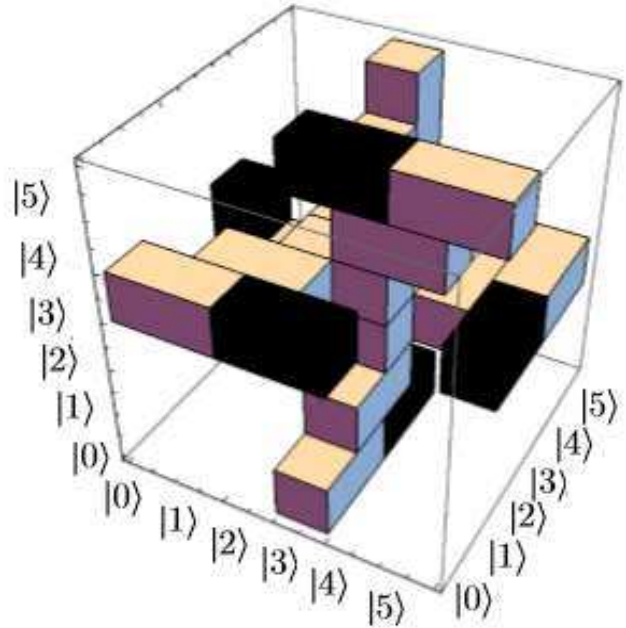}
	\caption{Product states representation in ${\mathbb{C}}^{6}\bigotimes{\mathbb{C}}^{6}\bigotimes{\mathbb{C}}^{6}$. Here, we represent a quantum state $\ket{i+\overline{i+1}
		}\ket{j}\ket{k}$ by a black cuboid and $\ket{i\pm\overline{i+1}
	}\ket{j}\ket{k}$ by a white cuboid where $\ket{i\pm\overline{i+1}
		}=\frac{1}{\sqrt{2}}(\ket{i}\pm\ket{i+1}),$ for integer $i$.
	}
\end{figure}

Now, we consider the following six states from the set, 
$$\begin{matrix}
      \ket{\psi_{0}}=\ket{0}\ket{3}\ket{2+3}, & 
      \ket{\psi_{1}}=\ket{1}\ket{3}\ket{1+2}\\
      \ket{\psi_{2}}=\ket{2}\ket{3}\ket{2+3}, & 
      \ket{\psi_{3}}=\ket{3}\ket{2+3}\ket{2}\\
      \ket{\psi_{4}}=\ket{4}\ket{3+4}\ket{2}, & \ket{\psi_{5}}=\ket{5}\ket{2+3}\ket{2}
 \end{matrix}$$\\
 As Alice's measurement preserves the  orthogonality of those product states, the post measurement states remain pairwise orthogonal. So, the above six states $\{\ket{\psi_{i}}\}_{i=0}^{5}$ also remain orthogonal. Therefore, we have\\
 $$\bra{\psi_{i}}{M_{A}}^{\dagger}M_{A}\otimes I \otimes I \ket{\psi_{j}}=0 ,\hspace{0.3 in}i\neq j$$
$$ \implies \bra{i}{M_{A}}^{\dagger}M_{A}\ket{j} \bra{\chi_{i}}\ket{\chi_{j}}=0, \hspace{0.3 in }i\neq j $$
$$\implies
m_{ij}=0 \hspace{0.3 in }i\neq j$$\\
Because $$\bra{\chi_{i}}\ket{\chi_{j}}=1$$ where \begin{align*}
      \ket{\chi_{i}}=\ket{3}\ket{2+3}, \ket{3}\ket{1+2}, \ket{3}\ket{2+3},\\
        \ket{2+3}\ket{2}, \ket{3+4}\ket{2}, \ket{2+3}\ket{2}
\end{align*}\\
Now we take, \\
$$\begin{matrix}
   \ket{\phi_{0}}=\ket{0+1}\ket{0}\ket{3}, & 
   \ket{\phi_{1}}=\ket{0-1}\ket{0}\ket{3} \\
   \ket{\phi_{2}}=\ket{1+2}\ket{1}\ket{3}, & 
   \ket{\phi_{3}}=\ket{1-2}\ket{1}\ket{3} \\
   \ket{\phi_{4}}=\ket{2+3}\ket{2}\ket{3}, & \ket{\phi_{5}}=\ket{2-3}\ket{2}\ket{3},\\ \ket{\phi_{6}}=\ket{3+4}\ket{2}\ket{4}, & 
   \ket{\phi_{7}}=\ket{3-4}\ket{2}\ket{4} \\
   \ket{\phi_{8}}=\ket{4+5}\ket{2}\ket{5}, & \ket{\phi_{9}}=\ket{4-5}\ket{2}\ket{5} 
\end{matrix}$$ \\
Since the matrix $M_A$ preserves the orthogonality of these states, thus we have\\
$$\bra{\phi_{0}}{M_{A}}^{\dagger}M_{A}\otimes I \otimes I \ket{\phi_{1}}=0$$
$$\implies\bra{0+1} {M_{A}}^{\dagger}M_{A}\ket{0-1}\bra{0}\ket{0}\bra{3}\ket{3}=0$$
$$\implies m_{00}-m_{01}+m_{10}-m_{11}=0$$
$$\implies m_{00}=m_{11}$$ 
since, $\ m_{ij}=0, i\neq j$
Similarly, we get $m_{11}=m_{22}$ by taking $\ket{\phi_{3}}$ and $\ket{\phi_{4}}$. Also, we get $ m_{22}=m_{33}$, $ m_{33}=m_{44}$,  $m_{44}=m_{55}$. Hence, 
$${M_{A}}^{\dagger}M_{A}=\begin{pmatrix}
a & 0 & . & . & . & 0\\
0 & a & . & . & . & 0\\
. & . & . &   &  & .\\
. & . &   & . &  & .\\
. & . &   &   & . & .\\
0 & 0 & .&.&.& a
\end{pmatrix}$$
for some $a\in \mathbb{R}$.\\
Since, $M_{A}$ is any arbitrary matrix that preserves orthogonality of the above product states. Thus for any measurement $\{{{M_{Ai}}^{\dagger}M_{Ai}}\}$ Alice can do only the trivial one. This implies the fact that the set of product states is LOCC indistinguishable.\hspace{1.6in} $\blacksquare$ \\\\
Theorem 1: In  ${\mathbb{C}}^{2d}\bigotimes{\mathbb{C}}^{2d}\bigotimes{\mathbb{C}}^{2d}$, where $d$ is odd, the set of $18(d-1)$ orthogonal product states\\\\
$\ket{\phi_{i+1}^\pm}=\ket{i\pm\overline{i+1}}\ket{i}\ket{d}$, $i=0,1,2,...,d-1.$\\
$\ket{\phi_{d+1+i}^\pm}=\ket{d}\ket{i\pm\overline{i+1}}\ket{i}$, $i=0,1,2,...,d-1.$\\
$\ket{\phi_{2d+1+i}^\pm}=\ket{i}\ket{d}\ket{i\pm\overline{i+1}}$, $i=0,1,2,...,d-1.$\\
$\ket{\phi_{2d+1+i}^\pm}=\ket{i\pm\overline{i+1}}\ket{d-1}\ket{i+1}$, $i=d,d+1,d+2,...,2d-2.$\\
$\ket{\phi_{3d+i}^\pm}=\ket{i+1}\ket{i\pm\overline{i+1}}\ket{d-1}$, $i=d,d+1,d+2,...,2d-2.$\\
$\ket{\phi_{4d-1+i}^\pm}=\ket{d-1}\ket{i+1}\ket{i\pm\overline{i+1}}$, $i=d,d+1,d+2,...,2d-2.$\\
$\ket{\phi_{6d-2+\frac{i}{2}}}=\ket{\overline{d-1}+d}\ket{i}\ket{d}$, $i=0,2,4,...,d-3.$\\
$\ket{\phi_{6d-1+\frac{d-3}{2}+\frac{i}{2}}}=\ket{d}\ket{\overline{d-1}+d}\ket{i}$, $i=0,2,4,...,d-3.$\\
$\ket{\phi_{7d-3+\frac{i}{2}}}=\ket{i}\ket{d}\ket{\overline{d-1}+d}$, $i=0,2,4,...,d-3.$\\
$\ket{\psi_{7d-2+\frac{d-3}{2}+\frac{i-1}{2}}}=\ket{\overline{d-2}+\overline{d-1}}\ket{i}\ket{d}$, $i=1,3,5,...,d-4.$\\
$\ket{\psi_{8d-5+\frac{i-1}{2}}}=\ket{d}\ket{\overline{d-2}+\overline{d-1}}\ket{i}$, $i=1,3,5,...,d-4.$\\
$\ket{\psi_{8d-5+\frac{d-3}{2}+\frac{i-1}{2}}}=\ket{i}\ket{d}\ket{\overline{d-2}+\overline{d-1}}$, $i=1,3,5,...,d-4.$\\
$\ket{\psi_{9d-9+\frac{i-d}{2}}}=\ket{\overline{d-1}+d}\ket{d-1}\ket{i}$, $i=d+2,d+4,...,2d-1.$\\
$\ket{\psi_{9d-9+\frac{i-1}{2}}}=\ket{i}\ket{\overline{d-1}+d}\ket{d-1}$, $i=d+2,d+4,...,2d-1.$\\
$\ket{\psi_{10d-10+\frac{i-d}{2}}}=\ket{d-1}\ket{i}\ket{\overline{d-1}+d}$, $i=d+2,d+4,...,2d-1.$\\
$\ket{\psi_{10d-11+\frac{i}{2}}}=\ket{d+\overline{d+1}}\ket{d-1}\ket{i}$, $i=d+3,d+5,...,2d-2.$\\
$\ket{\psi_{11d-12+\frac{i-d-1}{2}}}=\ket{i}\ket{d+\overline{d+1}}\ket{d-1}$, $i=d+3,d+5,...,2d-2.$\\ $\ket{\psi_{11d-14+\frac{i}{2}}}=\ket{d-1}\ket{i}\ket{d+\overline{d-1}}$, $i=d+3,d+5,...,2d-2.$\\\\
depicted in FIG. 2 are LOCC indistinguishable.\\\\
\textit{Proof}: See the Appendix A for explicit description of the proof. However, it is
easy to check that if three parties come together then the basis can be perfectly distinguished by LOCC. Furthermore, because of the construction, if all parties stand alone then no state can be eliminated by orthogonality preserving LOCC\cite{8,19}. Therefore, a resource state distributed among three parties is necessary to distinguish the basis\cite{22}. \\
\begin{figure}[h!]
	\centering
	\includegraphics[width=0.50\textwidth]{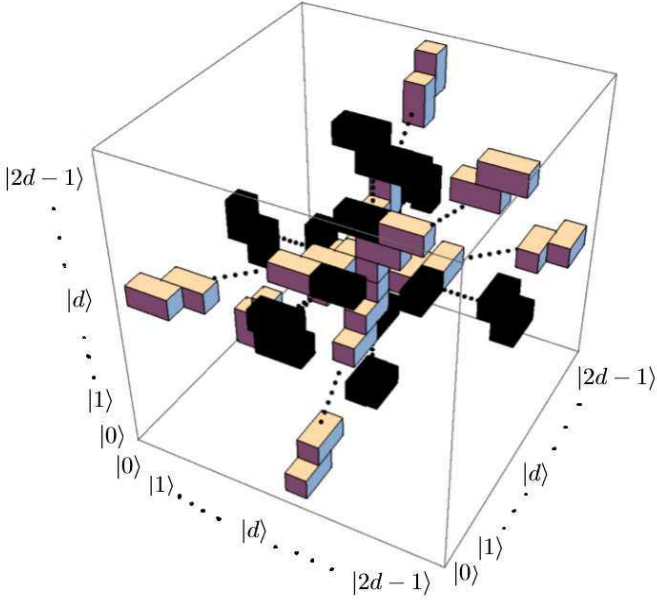}
	\caption{Product states representation in ${\mathbb{C}}^{2d}\bigotimes{\mathbb{C}}^{2d}\bigotimes{\mathbb{C}}^{2d}$, $d$ is odd. Here, we represent a quantum state $\ket{i+\overline{i+1}
		}\ket{j}\ket{k}$ by a black cuboid and $\ket{i\pm\overline{i+1}
		}\ket{j}\ket{k}$ by a white cuboid where $\ket{i\pm\overline{i+1}
		}=\frac{1}{\sqrt{2}}(\ket{i}\pm\ket{i+1}),$ for integer $i$.}
\end{figure}

Theorem 2: In  ${\mathbb{C}}^{2d}\bigotimes{\mathbb{C}}^{2d}\bigotimes{\mathbb{C}}^{2d}$, where $d$ is even, the set of $18(d-1)$ orthogonal product states\\\\
$\ket{\phi_{i+1}^\pm}=\ket{i\pm\overline{i+1}}\ket{i}\ket{d}$, $i=0,1,2,...,d-1.$\\
$\ket{\phi_{d+1+i}^\pm}=\ket{d}\ket{i\pm\overline{i+1}}\ket{i}$, $i=0,1,2,...,d-1.$\\
$\ket{\phi_{2d+1+i}^\pm}=\ket{i}\ket{d}\ket{i\pm\overline{i+1}}$, $i=0,1,2,...,d-1.$\\
$\ket{\phi_{2d+1+i}^\pm}=\ket{i\pm\overline{i+1}}\ket{d-1}\ket{i+1}$, $i=d,d+1,d+2,...,2d-2.$\\
$\ket{\phi_{3d+i}^\pm}=\ket{i+1}\ket{i\pm\overline{i+1}}\ket{d-1}$, $i=d,d+1,d+2,...,2d-2.$\\
$\ket{\phi_{4d-1+i}^\pm}=\ket{d-1}\ket{i+1}\ket{i\pm\overline{i+1}}$, $i=d,d+1,d+2,...,2d-2.$\\
$\ket{\phi_{6d-2+\frac{i-1}{2}}}=\ket{\overline{d-1}+d}\ket{i}\ket{d}$, $i=1,3,5,...,d-3.$\\
$\ket{\phi_{6d-1+\frac{d-4}{2}+\frac{i-1}{2}}}=\ket{d}\ket{\overline{d-1}+d}\ket{i}$, $i=1,3,5,...,d-3.$\\
$\ket{\phi_{7d-4+\frac{i-1}{2}}}=\ket{i}\ket{d}\ket{\overline{d-1}+d}$, $i=1,3,5,...,d-3.$\\
$\ket{\psi_{7d-3+\frac{d-4}{2}+\frac{i}{2}}}=\ket{\overline{d-2}+\overline{d-1}}\ket{i}\ket{d}$, $i=0,2,4,...,d-4.$\\
$\ket{\psi_{8d-6+\frac{i}{2}}}=\ket{d}\ket{\overline{d-2}+\overline{d-1}}\ket{i}$, $i=0,2,4,...,d-4.$\\
$\ket{\psi_{8d-5+\frac{d-4}{2}+\frac{i}{2}}}=\ket{i}\ket{d}\ket{\overline{d-2}+\overline{d-1}}$, $i=0,2,4,...,d-4.$\\
$\ket{\psi_{9d-9+\frac{i-d}{2}}}=\ket{\overline{d-1}+d}\ket{d-1}\ket{i}$, $i=d+2,d+4,...,2d-2.$\\
$\ket{\psi_{9d-10+\frac{i}{2}}}=\ket{i}\ket{\overline{d-1}+d}\ket{d-1}$, $i=d+2,d+4,...,2d-2.$\\
$\ket{\psi_{10d-11+\frac{i-d}{2}}}=\ket{d-1}\ket{i}\ket{\overline{d-1}+d}$, $i=d+2,d+4,...,2d-2.$\\
$\ket{\psi_{10d-11+\frac{i-3}{2}}}=\ket{d+\overline{d+1}}\ket{d-1}\ket{i}$, $i=d+3,d+5,...,2d-1.$\\
$\ket{\psi_{11d-13+\frac{i-d-1}{2}}}=\ket{i}\ket{d+\overline{d+1}}\ket{d-1}$, $i=d+3,d+5,...,2d-1.$\\ $\ket{\psi_{11d-13+\frac{i-3}{2}}}=\ket{d-1}\ket{i}\ket{d+\overline{d+1}}$, $i=d+3,d+5,...,2d-1.$\\\\	
depicted in FIG. 3 are LOCC indistinguishable.\\\\
\textit{Proof}: See the Appendix B for explicit description of the proof. At this point it is important to note that the local indistinguishability arises due to the twisted states, which are nothing but the linear superposition of the basis states\cite{5}. Thus it is clear that quantum superposition principle plays an important role in the phenomenon of quantum nonlocality without entanglement.\\
  \begin{figure}[h!]
	\centering
	\includegraphics[width=0.50\textwidth]{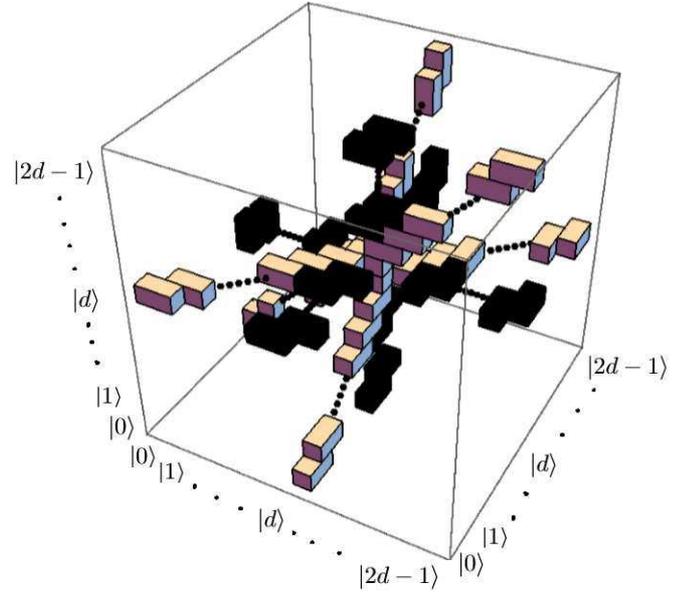}
	\caption{Product states representation in ${\mathbb{C}}^{2d}\bigotimes{\mathbb{C}}^{2d}\bigotimes{\mathbb{C}}^{2d}$, $d$ is even. Here, we represent a quantum state $\ket{i+\overline{i+1}
		}\ket{j}\ket{k}$ by a black cuboid and $\ket{i\pm\overline{i+1}
		}\ket{j}\ket{k}$ by a white cuboid where $\ket{i\pm\overline{i+1}
		}=\frac{1}{\sqrt{2}}(\ket{i}\pm\ket{i+1}),$ for integer $i$.}
\end{figure}\\
Now we consider the generalization in ${\mathbb{C}}^{k}\bigotimes{\mathbb{C}}^{l}\bigotimes{\mathbb{C}}^{m}$. For this we have to consider four cases. For better understanding, we start with some examples. \\
	Example 2: In
	${\mathbb{C}}^{5}\bigotimes{\mathbb{C}}^{5}\bigotimes{\mathbb{C}}^{5}$ the set of 31 orthogonal product states\\ $\ket{\phi_{1,2}}=\ket{1\pm2}\ket{4}\ket{2}$\\
	$\ket{\phi_{3,4}}=\ket{3\pm4}\ket{4}\ket{2}$\\
	$\ket{\phi_{5,6}}=\ket{4}\ket{2}\ket{1\pm2}$\\
	$\ket{\phi_{7,8}}=\ket{4}\ket{2}\ket{3\pm4}$\\
	$\ket{\phi_{9,10}}=\ket{2}\ket{1\pm2}\ket{4}$\\
	$\ket{\phi_{11,12}}=\ket{2}\ket{3\pm4}\ket{4}$\\
	$\ket{\phi_{13,14}}=\ket{2\pm3}\ket{0}\ket{2}$\\
	$\ket{\phi_{15,16}}=\ket{0\pm1}\ket{0}\ket{2}$\\
	$\ket{\phi_{17,18}}=\ket{2}\ket{2\pm3}\ket{0}$\\
	$\ket{\phi_{19,20}}=\ket{2}\ket{0\pm1}\ket{0}$\\
	$\ket{\phi_{21,22}}=\ket{0}\ket{2}\ket{2\pm3}$\\
	$\ket{\phi_{23,24}}=\ket{0}\ket{2}\ket{0\pm1}$\\
	$\ket{\phi_{25}}=\ket{1}\ket{2}\ket{2}$\\
	$\ket{\phi_{26}}=\ket{2}\ket{2}\ket{2}$\\
	$\ket{\phi_{27}}=\ket{3}\ket{2}\ket{2}$\\
	$\ket{\phi_{28}}=\ket{2}\ket{1}\ket{2}$\\
	$\ket{\phi_{29}}=\ket{2}\ket{3}\ket{2}$\\
	$\ket{\phi_{30}}=\ket{2}\ket{2}\ket{1}$\\
	$\ket{\phi_{31}}=\ket{2}\ket{2}\ket{3}$\\
	cannot be perfectly distinguished by LOCC.\\
	\textit{Proof}: See the Appendix C for the explicit description of the proof.\\
	\begin{figure}[h!]
		\centering
		\includegraphics[width=0.45\textwidth]{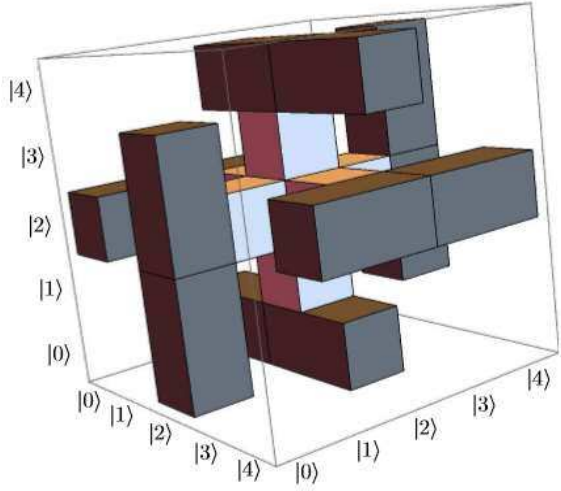}
		\caption{Product states representation in ${\mathbb{C}}^{5}\bigotimes{\mathbb{C}}^{5}\bigotimes{\mathbb{C}}^{5}$. Here, we represent a quantum state $\ket{i}\ket{j}\ket{k}$ by a white cuboid and $\ket{i\pm\overline{i+1}
			}\ket{j}\ket{k}$ by a black cuboid where $\ket{i\pm\overline{i+1}
			}=\frac{1}{\sqrt{2}}(\ket{i}\pm\ket{i+1}),$ for integer $i$.}
	\end{figure}
	In the following, we will show the local indistinguishability of this class of states in higher dimensions.\\\\
	Theorem 3: In  ${\mathbb{C}}^{2k+1}\bigotimes{\mathbb{C}}^{2l+1}\bigotimes{\mathbb{C}}^{2m+1}$, the set of $6(k+l+m)-5$ orthogonal product states\\\\
	$\ket{\phi_{i,i+1}}=\ket{i\pm\overline{i+1}}\ket{2l}\ket{m}$, $i=1,3,5,...,(2k-1).$\\
	$\ket{\phi_{2k+i,2k+i+1}}=\ket{2k}\ket{l}\ket{i\pm\overline{i+1}}$, $i=1,3,5,...,(2m-1).$\\
	$\ket{\phi_{2k+2m+i,2k+2m+i+1}}=\ket{k}\ket{i\pm\overline{i+1}}\ket{2m}$, $i=1,3,5,...,(2l-1).$\\
	$\ket{\phi_{2k+2m+2l+i-1,2k+2m+2l+i}}=\ket{i\pm\overline{i+1}}\ket{0}\ket{m}$, $i=2,4,6,...,(2k-2).$\\
	$\ket{\phi_{4k+2m+2l-1,4k+2m+2l}}=\ket{0\pm1}\ket{0}\ket{m}$\\
	$\ket{\phi_{4k+2m+2l+i-1,4k+2m+2l+i}}=\ket{k}\ket{i\pm\overline{i+1}}\ket{0}$, $i=2,4,6,...,(2l-2).$\\
	$\ket{\phi_{4k+4l+2m-1,4k+4l+2m}}=\ket{k}\ket{0\pm1}\ket{0}$\\
	$\ket{\phi_{4k+4l+2m+i-1,4k+4l+2m+i}}=\ket{0}\ket{l}\ket{i\pm\overline{i+1}}$, $i=2,4,6,...,(2m-2).$\\
	$\ket{\phi_{4k+4l+4m-1,4k+4l+4m}}=\ket{0}\ket{l}\ket{0\pm1}$\\
	$\ket{\phi_{4k+4l+4m+i}}=\ket{i}\ket{l}\ket{m}$, $i=1,2,3,...(2k-1).$\\
	$\ket{\phi_{6k+4l+4m+i-1}}=\ket{k}\ket{i}\ket{m}$, $i=1,2,3,...(l-1).$\\
	$\ket{\phi_{6k+4l+4m+i-2}}=\ket{k}\ket{i}\ket{m}$, $i=(l+1),(l+2),(l+3),...(2l-1).$\\
	$\ket{\phi_{6k+6l+4m+i-3}}=\ket{k}\ket{l}\ket{i}$, $i=1,2,3,...(m-1).$\\
	$\ket{\phi_{6k+6l+4m+i-4}}=\ket{k}\ket{l}\ket{i}$, $i=(m+1),(m+2),(m+3),...(2m-1).$\\
	cannot be perfectly distinguished by LOCC.\\\\
	\textit{Proof}. Proof is similar  as discussed above in Theorem 1 and Theorem 2.\\
	In Ref. \cite{8}, author constructed various kind of LOCC indistinguishable sets in bipartite systems. Here, we have extended it non trivially in tripartite systems. Further, it was shown that a two qubit maximally entangle state is sufficient to distinguish each class of states constructed in \cite{12}. But here in tripartite cases, it could be shown that either a GHZ state or two copies of Bell state is sufficient to distinguish each class of states. For special case $k=2,l=2,m=2$ the set of states is depicted in FIG. 4. 
\\\\
		\begin{figure}[h!]
		\centering
		\includegraphics[width=0.45\textwidth]{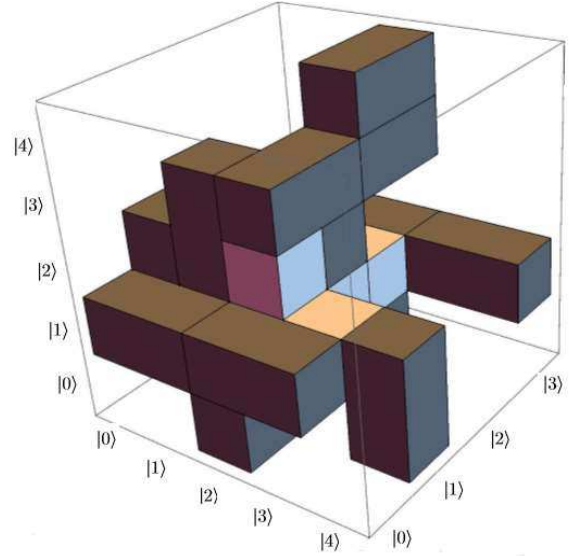}
		\caption{Product states representation in ${\mathbb{C}}^{5}\bigotimes{\mathbb{C}}^{5}\bigotimes{\mathbb{C}}^{4}$. Here, we represent a quantum state $\ket{i}\ket{j}\ket{k}$ by a white cuboid and $\ket{i\pm\overline{i+1}
			}\ket{j}\ket{k}$ by a black cuboid where $\ket{i\pm\overline{i+1}
			}=\frac{1}{\sqrt{2}}(\ket{i}\pm\ket{i+1}),$ for integer $i$}
	\end{figure}

	Theorem 4: In  ${\mathbb{C}}^{2k+1}\bigotimes{\mathbb{C}}^{2l+1}\bigotimes{\mathbb{C}}^{2m}$, the set of $6(k+l+m)-8$ orthogonal product states\\\\
	$\ket{\phi_{i,i+1}}=\ket{i\pm\overline{i+1}}\ket{2l}\ket{m-1}$, $i=1,3,5,...,(2k-1).$\\
	$\ket{\phi_{2k+1+i,2k+2+i}}=\ket{i\pm\overline{i+1}}\ket{0}\ket{m-1}$, $i=0,2,4,...,(2k-2).$\\
	$\ket{\phi_{4k+i,4k+1+i}}=\ket{2}\ket{i\pm\overline{i+1}}\ket{2m-1}$, $i=1,3,5,...,(2l-1).$\\
	$\ket{\phi_{4k+2l+1+i,4k+2l+2+i}}=\ket{2}\ket{i\pm\overline{i+1}}\ket{0}$, $i=0,2,4,...,(2l-2).$\\
	$\ket{\phi_{4k+4l+1+i,4k+4l+2+i}}=\ket{2k}\ket{1}\ket{i\pm\overline{i+1}}$, $i=0,2,4,...,(2m-4).$\\
	$\ket{\phi_{4k+4l+2m-2+i,4k+4l+2m-1+i}}=\ket{0}\ket{1}\ket{i\pm\overline{i+1}}$, $i=1,3,5,...,(2m-3).$\\
	$\ket{\phi_{4k+4l+4m-3,4k+4l+4m-2}}=\ket{1}\ket{1}\ket{\overline{2m-2}\pm\overline{2m-1}}$\\
	$\ket{\phi_{4k+4l+4m-2+i}}=\ket{i}\ket{1}\ket{m-1}$, $i=1,2,3,...,(2k-1).$\\ 
	$\ket{\phi_{6k+4l+4m-4+i}}=\ket{2}\ket{i}\ket{m-1}$, $i=2,3,4,...,(2l-1).$\\ 
	$\ket{\phi_{6k+6l+4m-5+i}}=\ket{2}\ket{1}\ket{i}$, $i=1,2,3,...,(m-2).$\\
	$\ket{\phi_{6k+6l+4m-6+i}}=\ket{2}\ket{1}\ket{i}$, $i=m,(m+1),(m+2)...,(2m-2).$\\
	cannot be perfectly distinguished by LOCC.\\\\
\textit{Proof:} Proof is similar to the cases discussed above in Theorem 1 and Theorem 2.\\
The idea of proof is similar with the method  presented by Wang et al.\cite{8}. We study the local indistinguishability of these
quantum states and also it can be shown that either a GHZ state or 2-copies of Bell states is/are sufficient to distinguish the above class of states. The set of states depicted in FIG. 5 represents the special case for $k=2,l=2,m=2$ of Theorem 4.
	\\\\

Theorem 5: In  ${\mathbb{C}}^{2k+1}\bigotimes{\mathbb{C}}^{2l}\bigotimes{\mathbb{C}}^{2m}$, the set of $6(k+l+m)-11$ orthogonal product states\\\\
$\ket{\phi_{i,i+1}}=\ket{i\pm\overline{i+1}}\ket{2l-1}\ket{m-1}$, $i=1,3,5,...,(2k-1).$\\
$\ket{\phi_{2k+1+i,2k+2+i}}=\ket{i\pm\overline{i+1}}\ket{0}\ket{m-1}$, $i=0,2,4,...,(2k-2).$\\
$\ket{\phi_{4k+i,4k+1+i}}=\ket{2}\ket{i\pm\overline{i+1}}\ket{2m-1}$, $i=1,3,5,...,(2l-3).$\\
$\ket{\phi_{4k+2l-1+i,4k+2l+i}}=\ket{2}\ket{i\pm\overline{i+1}}\ket{0}$, $i=0,2,4,...,(2l-4).$\\
$\ket{\phi_{4k+4l-3,4k+4l-2}}=\ket{2}\ket{\overline{2l-2}\pm\overline{2l-1}}\ket{2m-2}$\\
$\ket{\phi_{4k+4l-1+i,4k+4l+i}}=\ket{2k}\ket{1}\ket{i\pm\overline{i+1}}$, $i=0,2,4,...,(2m-4).$\\
$\ket{\phi_{4k+4l+2m-4+i,4k+4l+2m-3+i}}=\ket{0}\ket{1}\ket{i\pm\overline{i+1}}$, $i=1,3,5,...,(2m-3).$\\
$\ket{\phi_{4k+4l+4m-5,4k+4l+4m-4}}=\ket{1}\ket{1}\ket{\overline{2m-2}\pm\overline{2m-1}}$\\
$\ket{\phi_{4k+4l+4m-4+i}}=\ket{i}\ket{1}\ket{m-1}$, $i=1,2,3,...,(2k-1).$\\ 
$\ket{\phi_{6k+4l+4m-6+i}}=\ket{2}\ket{i}\ket{m-1}$, $i=2,3,4,...,(2l-2).$\\ 
$\ket{\phi_{6k+6l+4m-8+i}}=\ket{2}\ket{1}\ket{i}$, $i=1,2,3,...,(m-2).$\\
$\ket{\phi_{6k+6l+4m-9+i}}=\ket{2}\ket{1}\ket{i}$, $i=m,(m+1),(m+2)...,(2m-2).$\\
cannot be perfectly distinguished by LOCC.\\\\
\textit{Proof:} Proof is similar  as discussed above in Theorem 1 and Theorem 2.
\\\\
	
	Theorem 6: In  ${\mathbb{C}}^{2k}\bigotimes{\mathbb{C}}^{2l}\bigotimes{\mathbb{C}}^{2m}$, the set of $6(k+l+m)-14$ orthogonal product states\\\\
	$\ket{\phi_{i+1,i+2}}=\ket{i\pm\overline{i+1}}\ket{2l-1}\ket{2}$, $i=0,2,4,...,(2k-4).$\\
	$\ket{\phi_{2k-2+i,2k-1+i}}=\ket{i\pm\overline{i+1}}\ket{0}\ket{2}$, $i=1,3,5,...,(2k-5).$\\
	$\ket{\phi_{4k-5,4k-4}}=\ket{\overline{2k-2}\pm\overline{2k-1}}\ket{0}\ket{2}$\\
	$\ket{\phi_{4k-3+i,4k-2+i}}=\ket{2}\ket{i\pm\overline{i+1}}\ket{2m-1}$, $i=0,2,4,...,(2l-4).$\\
	$\ket{\phi_{4k+2l-6+i,4k+2l-5+i}}=\ket{2}\ket{i\pm\overline{i+1}}\ket{0}$, $i=1,3,5,...,(2l-5).$\\
	$\ket{\phi_{4k+4l-9,4k+4l-8}}=\ket{2}\ket{\overline{2l-2}\pm\overline{2l-1}}\ket{0}$\\
	$\ket{\phi_{4k+4l-7+i,4k+4l-6+i}}=\ket{2k-1}\ket{2}\ket{i\pm\overline{i+1}}$, $i=0,2,4,...,(2m-4).$\\
	$\ket{\phi_{4k+4l+2m-10+i,4k+4l+2m-9+i}}=\ket{0}\ket{2}\ket{i\pm\overline{i+1}}$, $i=1,3,5,...,(2m-5).$\\
	$\ket{\phi_{4k+4l+4m-13,4k+4l+4m-12}}=\ket{0}\ket{2}\ket{\overline{2m-2}\pm\overline{2m-1}}$\\
	$\ket{\phi_{4k+4l+4m-11,4k+4l+4m-10}}=\ket{\overline{2k-3}\pm\overline{2k-2}}\ket{1}\ket{2}$\\
	$\ket{\phi_{4k+4l+4m-9,4k+4l+4m-8}}=\ket{2}\ket{\overline{2l-3}\pm\overline{2l-2}}\ket{1}$\\
	$\ket{\phi_{4k+4l+4m-7,4k+4l+4m-6}}=\ket{1}\ket{2}\ket{\overline{2m-3}\pm\overline{2m-2}}$\\
	$\ket{\phi_{4k+4l+4m-7+i}}=\ket{2}\ket{2}\ket{i}$, $i=2,3,...,(2m-2).$\\ 
	$\ket{\phi_{4k+4l+6m-11+i}}=\ket{i}\ket{2}\ket{2}$, $i=3,4,...,(2k-2).$\\ 
	$\ket{\phi_{6k+4l+6m-15+i}}=\ket{2}\ket{i}\ket{2}$, $i=3,4,...,(2l-2).$\\
	$\ket{\phi_{6k+6l+6m-16}}=\ket{2}\ket{2}\ket{1}$\\
	$\ket{\phi_{6k+6l+6m-15}}=\ket{1}\ket{2}\ket{2}$\\
	$\ket{\phi_{6k+6l+6m-14}}=\ket{2}\ket{1}\ket{2}$\\
	cannot be perfectly distinguished by LOCC.\\\\
	\textit{Proof:} Proof is similar  as discussed above in Theorem 1 and Theorem 2. The set of states depicted in FIG. 6 represents the special case for $k=3,l=3,m=3$ of Theorem 6.
	\\\\
	\begin{figure}[h!]
	\centering
	\includegraphics[width=0.45\textwidth]{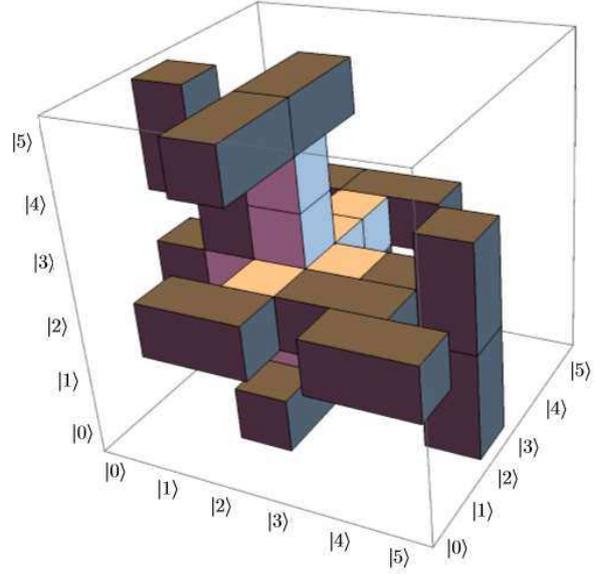}
	\caption{Product states representation in ${\mathbb{C}}^{6}\bigotimes{\mathbb{C}}^{6}\bigotimes{\mathbb{C}}^{6}$. Here, we represent a quantum state $\ket{i}\ket{j}\ket{k}$ by a white cuboid and $\ket{i\pm\overline{i+1}
		}\ket{j}\ket{k}$ by a black cuboid where $\ket{i\pm\overline{i+1}
		}=\frac{1}{\sqrt{2}}(\ket{i}\pm\ket{i+1}),$ for integer $i$}
\end{figure}	
	
One may observe that each class of product states constructed above are not an UPB. But the sets can be extended to orthonormal product bases. From the above discussion, it is also prominent that the sets of states cannot be perfectly distinguished by LOCC. This is because of the fact that for local distinguishability it is necessary to eliminate state(s) of a given set, which is not possible in the given scenario. The special about the sets(Th.1-Th.6) is that they are the LOCC indistinguishable sets containing minimum number of states with respect to their dimensions. Because of the symmetric structure of those sets it indicates that no one of the three parties cannot eliminate any state from those sets. Each state of those sets contributes to prove the measurements trivial for all three parties; i.e., all parties doing only trivial one and any state from those sets cannot be eliminated.

\section{Discrimination protocol}
Here, we will show that each of the above sets of LOCC indistinguishable product states in ${\mathbb{C}}^{k}\bigotimes{\mathbb{C}}^{l}\bigotimes{\mathbb{C}}^{m}$ can be distinguished if the parties share an auxiliary GHZ state $\ket{\psi}_{abc}=\frac{1}{\sqrt{2}}(\ket{000}+\ket{111})$. We will share this tripartite entanglement between parties by the technique described in \cite{21,22}. For better understanding, we will prove explicitly for lower dimensional one and the result hold automatically for higher dimensions also.\\\\
Example 3: In      ${\mathbb{C}}^{6}\bigotimes{\mathbb{C}}^{6}\bigotimes{\mathbb{C}}^{6}$ the set of 36 orthogonal product states\\
$$\begin{Bmatrix}
\begin{matrix}
\ket{\psi_{1}}=\ket{2+3}\ket{0}\ket{3}\\
\ket{\psi_{2}}=\ket{3}\ket{2+3}\ket{0}\\
\ket{\psi_{3}}=\ket{0}\ket{3}\ket{2+3}
\end{matrix} & 
\begin{matrix}
\ket{\psi_{4}}=\ket{2+3}\ket{2}\ket{5}\\
\ket{\psi_{5}}=\ket{5}\ket{2+3}\ket{2}\\
\ket{\psi_{6}}=\ket{2}\ket{5}\ket{2+3}
\end{matrix}\\\\
\begin{matrix}
\ket{\psi_{7,8}}=\ket{2\pm3}\ket{2}\ket{3}\\
\ket{\psi_{9,10}}=\ket{3}\ket{2\pm3}\ket{2}\\
\ket{\psi_{11,12}}=\ket{2}\ket{3}\ket{2\pm3}
\end{matrix} & 
\begin{matrix}
\ket{\psi_{13,14}}=\ket{3\pm4}\ket{2}\ket{4}\\
\ket{\psi_{15,16}}=\ket{4}\ket{3\pm4}\ket{2}\\
\ket{\psi_{17,18}}=\ket{2}\ket{4}\ket{3\pm4}
\end{matrix}\\\\ 
\begin{matrix} 
\ket{\psi_{19,20}}=\ket{1\pm2}\ket{1}\ket{3}\\
\ket{\psi_{21,22}}=\ket{3}\ket{1\pm2}\ket{1}\\
\ket{\psi_{23,24}}=\ket{1}\ket{3}\ket{1\pm2}
\end{matrix} & 
\begin{matrix}
\ket{\psi_{25,26}}=\ket{0\pm1}\ket{0}\ket{3}\\
\ket{\psi_{27,28}}=\ket{3}\ket{0\pm1}\ket{0}\\
\ket{\psi_{29,30}}=\ket{0}\ket{3}\ket{0\pm1}
\end{matrix}\\\\
\begin{matrix} 
\ket{\psi_{31,32}}=\ket{4\pm5}\ket{2}\ket{5}\\
\ket{\psi_{33,34}}=\ket{5}\ket{4\pm5}\ket{2}\\
\ket{\psi_{35,36}}=\ket{2}\ket{5}\ket{4\pm5}
\end{matrix}\\\\
\end{Bmatrix}$$\\ can be distinguished by LOCC with one-copy of GHZ state.\\\\\\
Proof: First of all, three parties, Alice, Bob and Charlie share a GHZ state $\ket{\psi}_{abc}=\frac{1}{\sqrt{2}}(\ket{000}+\ket{111})$. Charlie performs a two-outcome measurement where each outcome corresponds to a rank-6 projector:
\begin{multline*}
C_1=\ket{00}_{Cc}\bra{00}+\ket{10}_{Cc}\bra{10}+\ket{20}_{Cc}\bra{20}\\
+\ket{31}_{Cc}\bra{31}+\ket{41}_{Cc}\bra{41}+\ket{51}_{Cc}\bra{51}
\end{multline*}
\begin{multline}
$$C_2=\ket{01}_{Cc}\bra{01}+\ket{11}_{Cc}\bra{11}+\ket{21}_{Cc}\bra{21}\\
+\ket{30}_{Cc}\bra{30}+\ket{40}_{Cc}\bra{40}+\ket{50}_{Cc}\bra{50}$$
\end{multline}
  \begin{figure}[h!]
	\centering
	\includegraphics[width=0.45\textwidth]{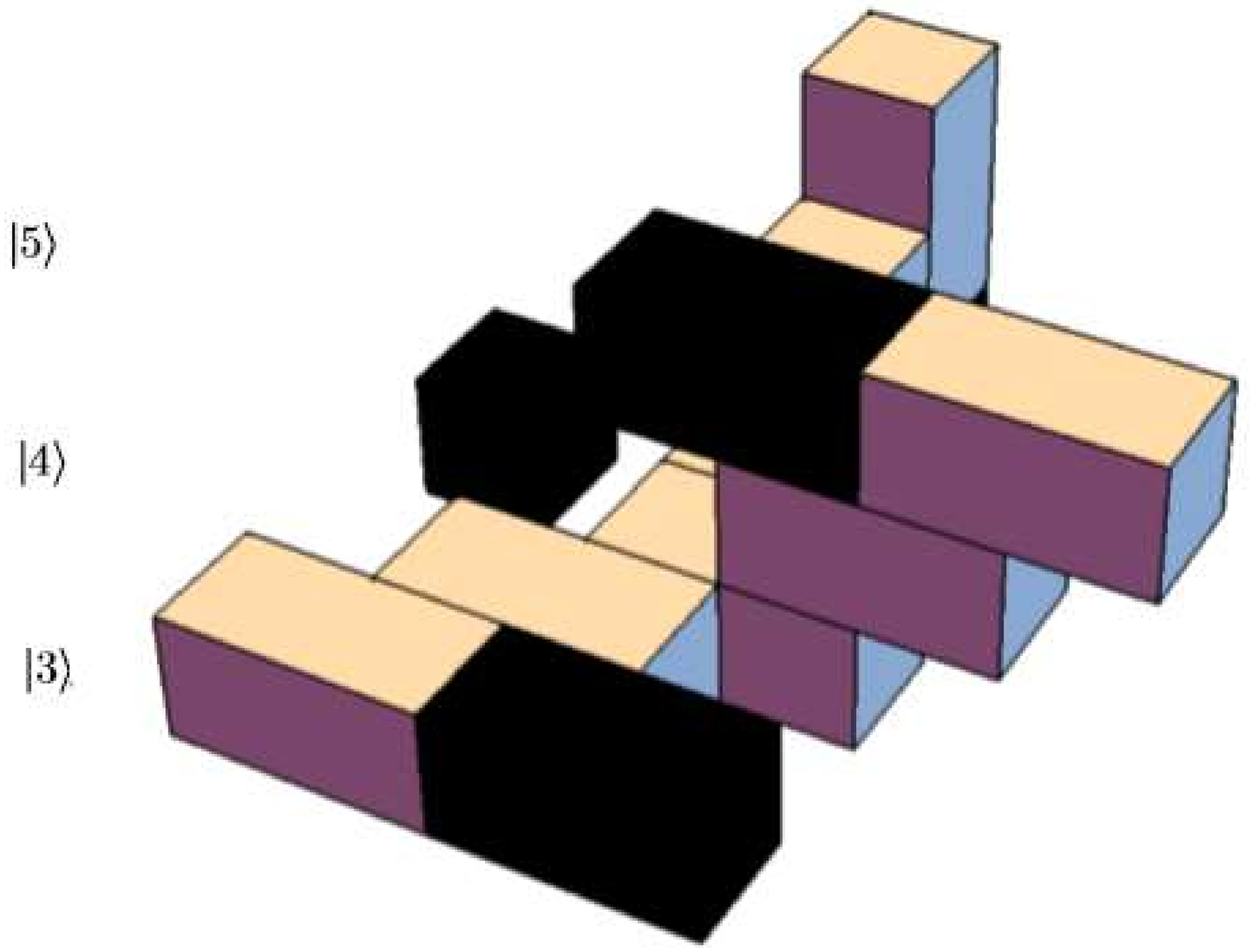}
	$$\ket{111}_{abc}$$
	\includegraphics[width=0.45\textwidth]{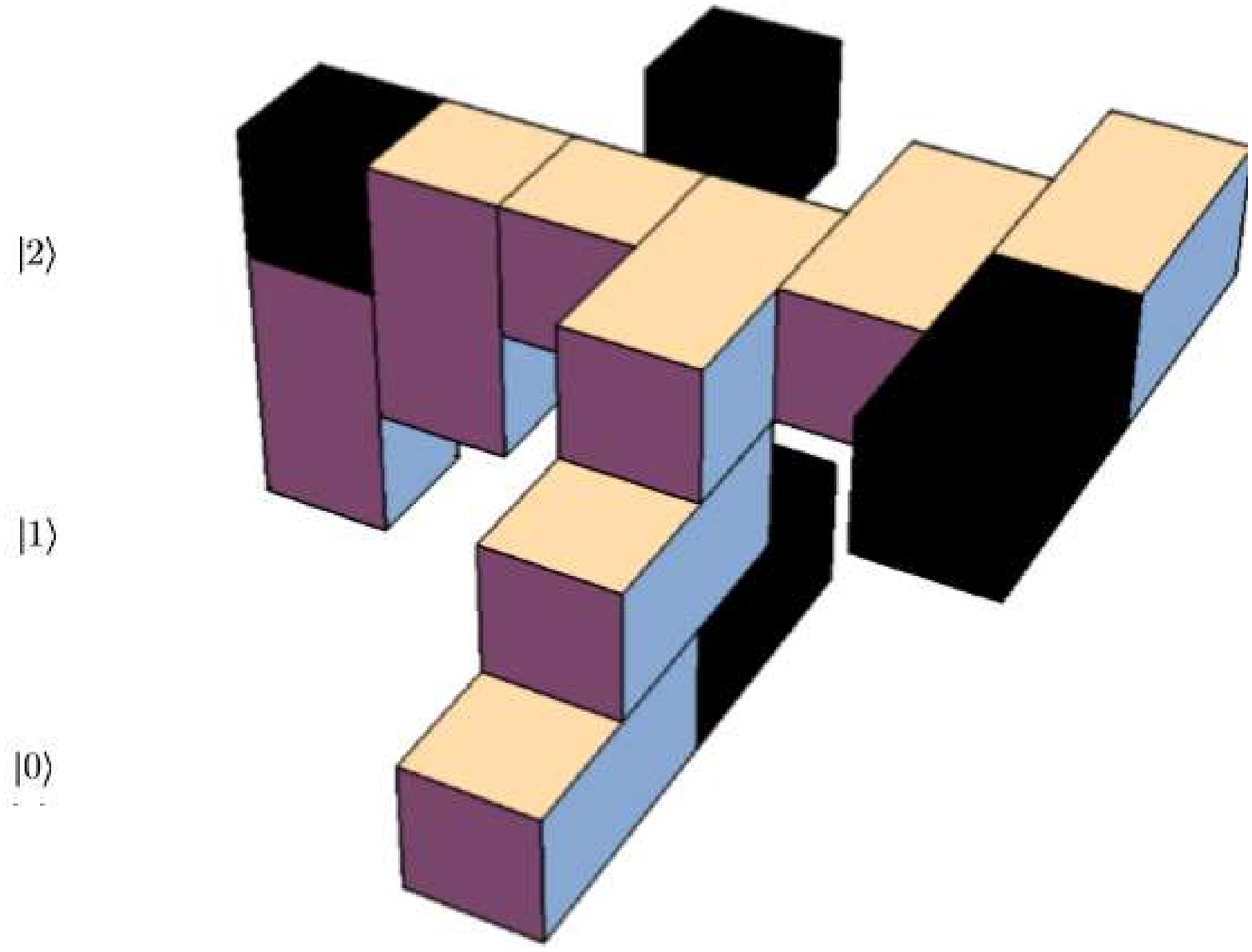}
	$$\ket{000}_{abc}$$
	\caption{Representation of product states after operating projector $C_1$. Here, we represent a quantum state $\ket{i+\overline{i+1}
		}\ket{j}\ket{k}$ by a black cuboid and $\ket{i\pm\overline{i+1}
		}\ket{j}\ket{k}$ by a gray cuboid where $\ket{i\pm\overline{i+1}
		}=\frac{1}{\sqrt{2}}(\ket{i}\pm\ket{i+1}),$ for integer $i$. }
\end{figure}\\
After operating the projector $C_1$ on systems $Cc$, each of the initial states
is transformed into\\
$$\begin{pmatrix}
\begin{Bmatrix}
\begin{matrix}
\ket{3}\ket{0\pm1}\ket{0}\\
\ket{0}\ket{3}\ket{0\pm1}\\ 
\ket{3}\ket{1\pm2}\ket{1}\\
\ket{1}\ket{3}\ket{1\pm2}\\
\ket{4}\ket{3\pm4}\ket{2}
\end{matrix} & 
\begin{matrix}
\ket{3}\ket{2\pm3}\ket{2}\\
\ket{5}\ket{4\pm5}\ket{2}\\
\ket{5}\ket{2+3}\ket{2}\\
\ket{3}\ket{2+3}\ket{0}
\end{matrix}
\end{Bmatrix}\bigotimes\ket{000}_{abc}$$\\
$$\begin{Bmatrix}
\begin{matrix}
\ket{0\pm1}\ket{0}\ket{3}\\
\ket{2+3}\ket{0}\ket{3}\\
\ket{4\pm5}\ket{2}\ket{5}\\ 
\ket{1\pm2}\ket{1}\ket{3}\\
\ket{2+3}\ket{2}\ket{5}
\end{matrix} & 
\begin{matrix}
\ket{2}\ket{4}\ket{3\pm4}\\
\ket{2\pm3}\ket{2}\ket{3}\\
\ket{3\pm4}\ket{2}\ket{4}\\
\ket{2}\ket{5}\ket{4\pm5}
\end{matrix}
\end{Bmatrix}\bigotimes\ket{111}_{abc}$$\\

$$\begin{Bmatrix}
\ket{032}_{ABC}\bigotimes\ket{000}_{abc}+\ket{033}_{ABC}\bigotimes\ket{111}_{abc}\\\ket{252}_{ABC}\bigotimes\ket{000}_{abc}+\ket{253}_{ABC}\bigotimes\ket{111}_{abc}\\
\ket{232}_{ABC}\bigotimes\ket{000}_{abc}\pm\ket{233}_{ABC}\bigotimes\ket{111}_{abc}
\end{Bmatrix}
\end{pmatrix}$$\\
i.e.,
\begin{multline}
\ket{{\psi\prime}_{i}}=\ket{\psi_{i}}_{ABC}\bigotimes\ket{000}_{abc},  i=2,5,9,10,15,16,\\
21,22,23,24,27,28,29,30,33,34\\
\ket{{\psi\prime}_{i}}=\ket{\psi_{i}}_{ABC}\bigotimes\ket{111}_{abc},  i=1,4,7,8,13,14,\\
17,18,19,20,25,26,31,32,35,36.\\
\ket{{\psi\prime}_{3}}=\ket{032}_{ABC}\bigotimes\ket{000}_{abc}+\ket{033}_{ABC}\bigotimes\ket{111}_{abc}\\
\ket{{\psi\prime}_{6}}=\ket{252}_{ABC}\bigotimes\ket{000}_{abc}+\ket{253}_{ABC}\bigotimes\ket{111}_{abc}\\
\ket{{\psi\prime}_{11,12}}=\ket{232}_{ABC}\bigotimes\ket{000}_{abc}\pm\ket{233}_{ABC}\bigotimes\ket{111}_{abc}
\end{multline}\\as shown in FIG. 7.
After operating the projector $C_2$ on systems $Cc$, it creates new states
which differ from the states $(2)$ only by ancillary systems
$\ket{000}_{abc}\rightarrow\ket{111}_{abc}$ and $\ket{111}_{abc}\rightarrow\ket{000}_{abc}$. Then the latter can be handled using the exact same method as for $C_1$. Thus, we only
need to discuss the behaviour after operating $C_1$.
Now we describe the method to distinguish those states how the parties act. After $C_1$ clicks, Bob makes  four-outcome projective measurement. The first outcome $B_1=\ket{0}_B\bra{0}\bigotimes\ket{1}_b\bra{1}$, which leaves $\ket{{\psi\prime}_{1,25,26}}$ and remain same for all other states in $(2)$. Then Alice can easily discriminate the three states by projecting onto $\ket{0\pm1}_A$ and $\ket{2+3}_A$. The second outcome is $B_2=\ket{1}_B\bra{1}\bigotimes\ket{1}_b\bra{1}$. The only remaining possibility is $\ket{{\psi\prime}_{19,20}}$, which can be successfully identified by Alice by projecting onto  $\ket{1\pm2}_A$. Using rank one projector $B_3=\ket{2}_B\bra{2}\bigotimes\ket{1}_b\bra{1}$ on Bob's Hilbert space leaves $\ket{{\psi\prime}_{4,7,8,13,14,31,32}}$ and annihilates other states. Corresponding to $B_3$ when Charlie uses projector $C_{31}=\ket{3}_C\bra{3}\bigotimes\ket{1}_c\bra{1}$, it leaves $\ket{{\psi\prime}_{7,8}}$, which can be easily distinguished by projectors  $\ket{2\pm3}_A$. When Charlie uses projector $C_{32}=\ket{4}_C\bra{4}\bigotimes\ket{1}_c\bra{1}$, it leaves $\ket{{\psi\prime}_{13,14}}$, which can be easily distinguished by projectors  $\ket{3\pm4}_A$. Also when Charlie uses projector $C_{33}=\ket{5}_C\bra{5}\bigotimes\ket{1}_c\bra{1}$, it leaves $\ket{{\psi\prime}_{4,31,32}}$, which can be easily distinguished by projectors  $\ket{2+3}_A$ and $\ket{4\pm5}_A$. \\\\
Bob's last outcome is $B_4=I_B-\left( B_1+B_2+B_3\right) $, which leaves all other states excepting $12$ states which has been discriminated above. Now corresponding to $B_4$, Alice makes $4$-outcome projective measurement. The first outcome $A_{41}=\ket{0}_A\bra{0}\bigotimes\left( \ket{0}_a\bra{0}+\ket{1}_a\bra{1}\right) $  which leaves $\ket{{\psi\prime}_{3,29,30}}$ and remain same for all other states. If Charlie projects onto $\ket{0\pm1}$, then $\ket{{\psi\prime}_{29,30}}$ are successfully identified and if projects onto other,  $\ket{{\psi\prime}_3}$ clicks. Now, if $A_{42}=\ket{0}_A\bra{0}\bigotimes \ket{0}_a\bra{0}$ occur, it leaves $\ket{{\psi\prime}_{23,24}}$ and thus successfully identified by Charlie by projecting onto $\ket{1\pm2}$. Alice's third outcome $A_{43}=\ket{2}_A\bra{2}\bigotimes\left( \ket{0}_a\bra{0}+\ket{1}_a\bra{1}\right) $  which leaves $\ket{{\psi\prime}_{6,11,12,17,18,35,36}}$ and remain same for all other states. Again, if Bob projects onto $\ket{3}_B\bra{3}\bigotimes\left( \ket{0}_b\bra{0}+\ket{1}_b\bra{1}\right) $, it leaves two states $\ket{{\psi\prime}_{11,12}}$ and thus successfully distinguished. If Bob projects onto $\ket{4}_B\bra{4}\bigotimes\ket{1}_b\bra{1}$ which leaves $\ket{{\psi\prime}_{17,18}}$ and thus Charlie will distinguish by projecting onto $\ket{3\pm4}_C$. Also, if Bob projects onto 
$\ket{5}_B\bra{5}\bigotimes\left( \ket{0}_b\bra{0}+\ket{1}_b\bra{1}\right) $ which leaves $\ket{{\psi\prime}_{6,35,36}}$ and thus Charlie distinguish $\ket{{\psi\prime}_{35,36}}$ by projecting onto $\ket{4\pm5}_C$ and $\ket{{\psi\prime}_{6}}$ by projecting onto other.\\\\
Alice's last outcome is $A_{44}=I_A-\left( A_{41}+A_{42}+A_{43}\right) $, which leaves $\ket{{\psi\prime}_{2,5,9,10,15,16,21,22,27,28,33,34}}$. Now corresponding to $A_{44}$ Charlie makes three outcome projective measurement. The first outcome $C_{441}=\ket{0}_C\bra{0}\bigotimes\ket{0}_c\bra{0}$ leaves $\ket{{\psi\prime}_{2,27,28}}$ and annihilates other states. Next Bob can distinguish those by projecting onto  $\ket{0\pm1}_B$ and $\ket{2+3}_B$. If the second outcome $C_{442}=\ket{1}_C\bra{1}\bigotimes\ket{0}_c\bra{0}$ occur, this eliminates $\ket{{\psi\prime}_{21,22}}$ and annihilates other states. Now Bob can distinguish those by projecting onto  $\ket{1\pm2}_B$. If Charlie's last outcome $C_{443}=\ket{2}_C\bra{2}\bigotimes\ket{0}_c\bra{0}$ occur, then it eliminates the remaining states $\ket{{\psi\prime}_{5,9,10,15,16,33,34}}$. Corresponding to $C_{443}$ Alice makes three outcome projective measurement. The first outcome  $A_{4431}=\ket{3}_A\bra{3}\bigotimes \ket{0}_a\bra{0}$, which eliminates $\ket{{\psi\prime}_{9,10}}$. Thus Bob can distinguish those by projecting onto $\ket{2\pm3}_B$. The second outcome $A_{4432}=\ket{4}_A\bra{4}\bigotimes \ket{0}_a\bra{0}$ eliminates $\ket{{\psi\prime}_{15,16}}$, which can be distinguished by  Bob using the projectors $\ket{3\pm4}_B$. The last outcome $A_{4433}=\ket{5}_A\bra{5}\bigotimes \ket{0}_a\bra{0}$ eliminates $\ket{{\psi\prime}_{5,33,34}}$, which can be distinguished by  Bob using the projectors $\ket{2+3}_B$ and $\ket{4\pm5}_B$. This completes our argument and proof of the result.\hspace{2.5in}$\blacksquare$\\
Interestingly, it can be shown that this class of states can be perfectly distinguished by LOCC if any two parties share a Bell state. But there is a restriction on that protocol, one of them who share the Bell state have to start the protocol. But, if a GHZ state is used as a resource then any one can start the discrimination protocol. See the Appendix D for Bell state discrimination.\\
Theorem 7:  In  ${\mathbb{C}}^{2d}\bigotimes{\mathbb{C}}^{2d}\bigotimes{\mathbb{C}}^{2d}$, where $d$ is odd or even, the set of $18(d-1)$ orthogonal product states can be distinguished by LOCC with one-copy of GHZ state.\\\\
Proof: The method is similar to the approach we have used previously for $d=3$. First of all, Alice, Bob and Charlie share a GHZ state $\ket{\psi}_{abc}=\frac{1}{\sqrt{2}}(\ket{000}+\ket{111})$. Then Charlie will perform a two-
outcome measurement, each outcome corresponding to a rank-2d projector:\\
\begin{multline*}
$$C_1=\ket{00}_{Cc}\bra{00}+\ket{10}_{Cc}\bra      {10}+\ket{20}_{Cc}\bra{20}+...\\
+\ket{\overline{d-1}0}_{Cc}\bra{\overline{d-1}0}\\
+\ket{d1}_{Cc}\bra{d1}+\ket{\overline{d+1}1}_{Cc}\bra{\overline{d+1}1}+...\\
+\ket{\overline{2d-1}1}_{Cc}\bra{\overline{2d-1}1}
\end{multline*}
\begin{multline}
C_2=\ket{01}_{Cc}\bra{01}+\ket{11}_{Cc}\bra{11}+\ket{21}_{Cc}\bra{21}+...\\
+\ket{\overline{d-1}1}_{Cc}\bra{\overline{d-1}1}\\
+\ket{d0}_{Cc}\bra{d0}+\ket{\overline{d+1}0}_{Cc}\bra{\overline{d+1}0}+...\\
+\ket{\overline{2d-1}0}_{Cc}\bra{\overline{2d-1}0}$$
\end{multline}
Alice, Bob and Charlie will do a sequence of measurements to distinguish those states as we have done in previous one.\\\\

Theorem 8: In  ${\mathbb{C}}^{2k+1}\bigotimes{\mathbb{C}}^{2l+1}\bigotimes{\mathbb{C}}^{2m+1}$, a GHZ state shared between three parties is sufficient to perfectly distinguished the set of $6(k+l+m)-5$ orthogonal product states by LOCC.\\
\textit{Proof:} First of all, Alice, Bob and Charlie share a GHZ state $\ket{\psi}_{abc}=\frac{1}{\sqrt{2}}(\ket{000}+\ket{111})$. Then Charlie will perform a two-
outcome measurement, each outcome corresponding to a rank-(2m-1) projector:\\
\begin{multline*}
$$C_1=\ket{00}_{Cc}\bra{00}+\ket{10}_{Cc}\bra      {10}+\ket{20}_{Cc}\bra{20}+.......\\+\ket{(2m-1)0}_{Cc}\bra{(2m-1)0}
+\ket{(2m)1}_{Cc}\bra{(2m)1}
\end{multline*}
\begin{multline*}
C_2=\ket{01}_{Cc}\bra{01}+\ket{11}_{Cc}\bra{11}+\ket{21}_{Cc}\bra{21}+......\\
+\ket{(2m-1)1}_{Cc}\bra{(2m-1)1}
+\ket{(2m)0}_{Cc}\bra{(2m)0}$$
\end{multline*}
After operating the projector $C_1$ on systems $Cc$, each of the initial states
will transform into\\ 
\begin{widetext}
	\begin{multline*}
	\ket{{\phi\prime}_{i}}=\ket{\phi_{i}}_{ABC}\bigotimes\ket{000}_{abc} ,  i=1,2,...,(2k+2m-2),(2k+2l+2m+1),...,(6k+6l+6m-5)\\
	\ket{{\phi\prime}_{i}}=\ket{\phi_{i}}_{ABC}\bigotimes\ket{111}_{abc} ,  i=(2k+2m+1),(2k+2m+2),...(2k+2l+2m)\\
	\ket{{\phi\prime}_{2k+2m-1,2k+2m}}=\ket{(2k)(l)(2m-1)}_{ABC}\bigotimes\ket{000}_{abc}+\ket{(2k)(l)(2m)}_{ABC}\bigotimes\ket{111}_{abc}\\
	\end{multline*}\\
\end{widetext}
If $C_1$ click, Bob makes  two-outcome projective measurement. The first outcome $B_1=\ket{2l}_B\bra{2l}\bigotimes\ket{0}_b\bra{0}$, which leaves $\ket{{\phi\prime}_{1,2,...,2k}}$ and those can be distinguished by projecting onto $Aa$ system. If $B_2=I-B_1$ click, it isolates $\ket{{\phi\prime}_{2k+1,2k+2,...6(k+l+m)-5}}$. After $B_2$ click, Alice makes  three-outcome projective measurement, $A_1=\ket{k}_A\bra{k}\bigotimes\ket{1}_a\bra{1}$, $A_2=\ket{2k}_A\bra{2k}\bigotimes\ket{0}_a\bra{0}+\ket{2k}_A\bra{2k}\bigotimes\ket{1}_a\bra{1}$, $A_3=I-A_1-A_2$. If $A_1$ clicks, it isolates $\ket{{\phi\prime}_{2k+2m+1,2k+2m+2,...2k+2l+2m}}$, which can be easily distinguished by projecting onto the $Bb$ system. If  $A_2$ clicks, it isolates $\ket{{\phi\prime}_{2k+1,2k+2,...2k+2m}}$, which can be easily distinguished by projecting onto $Cc$ system. If  $A_3$ clicks, it isolates $\ket{{\phi\prime}_{2k+2m+2l+1,2k+2m+2l+2,...,6(k+l+m)-5}}$. For $A_3$, Bob  makes two outcome measurement, 
$B_1=\ket{00}_{Bb}\bra{00}+ ....+\ket{(l-1)0}_{Bb}\bra{(l-1)0}$,  $B_2=\ket{l0}_{Bb}\bra{l0}+ ....+\ket{(2l-1)0}_{Bb}\bra{(2l-1)0}$.\\
If $B_1$ clicks it isolates $\ket{\phi\prime_{i}}$, $i=2k+2m+2l+1,...,4k+2m+2l,6k+4l+4m,...,6k+5l+4m-2$, which can be distinguished further. Similarly, if $B_2$ clicks, it can be distinguished as previously done.\\
That is, we have succeeded in designing a protocol to distinguish the states by LOCC with GHZ state. This completes the proof.\hspace{2.0in}$\blacksquare$\\
Similarly, by the protocol we have described in the previous theorem, we can show that a GHZ state is sufficient to distinguish each of the above classes in ${\mathbb{C}}^{2k+1}\bigotimes{\mathbb{C}}^{2l+1}\bigotimes{\mathbb{C}}^{2m}$, ${\mathbb{C}}^{2k+1}\bigotimes{\mathbb{C}}^{2l}\bigotimes{\mathbb{C}}^{2m}$,\\
${\mathbb{C}}^{2k}\bigotimes{\mathbb{C}}^{2l}\bigotimes{\mathbb{C}}^{2m}$.
But for these four classes of states (Th.3 - Th.6), it can be verified that at least 2-copy of Bell states needed to distinguish each of those classes. See the Appendix D for Bell state discrimination.\\\\
Since, all the sets are symmetric, it is really not important which pair of parties holds the resource state. From the above protocol it is also clear that the required entanglement resource to accomplish the task of distinguishing the non-local sets does not depend on the dimension of the subsystems. So, in the above protocol entanglement is used more efficiently than a teleportation based protocol. It is also to be noted that all the LOCC indistinguishable sets constructed in Th.1-Th.6 are minimal in cardinality. Also, the above classes of states can be distinguished by LOCC with the help of a GHZ state as a resource. In other words, it can say that those sets are minimal sets which are distinguishable through GHZ state. The protocol indicates the advantage of using GHZ state as a resource for discrimination some of those classes. For that, those classes are strong enough to show genuine entanglement more powerful.

\section{CONCLUSION AND OPEN PROBLEMS}
The phenomenon of quantum nonlocality without entanglement changes our sense of entanglement that leads to nonlocality. Most of the related results are about bipartite quantum system. So it is important to study multipartite systems where the complexity grows with the number of parties. Here we successfully done with the tripartite cases. In this work, we have constructed some new classes of orthogonal product states in tripartite systems which cannot be perfectly distinguished by LOCC. We are able to distinguish them with entanglement assisted local discrimination protocols. The protocols are resource efficient as they consume less entanglement than a teleportation based protocol. Moreover, we have discrimination protocols for those non-local classes of orthogonal product states with different configurations of entangled resources. Interestingly, we get advantage by using genuine entanglement as a resource for discrimination some of those classes. Here are two interesting open problems: The first is to find the LOCC indistinguishable sets of lesser number of states for arbitrarily high-dimensional multipartite quantum systems. Second, the question of optimality of the entangled resources used in our discrimination protocols remains open. In a given Hilbert space, the subsets constructed here are small sets. That is the number of states contained in the set is much lesser than the net dimension of the Hilbert space. Hence an essential search in this direction is to find out the number of orthogonal product states from which no one can be eliminated by OPPOVM. Explicit constructions of such sets are also important for any multipartite quantum system.  
\section*{ACKNOWLEDGEMENTS}
The authors I. Chattopadhyay and D. Sarkar acknowledge the work as part of QuST initiatives by DST India. The author A. Bhunia acknowledges the support from UGC, India

\section*{Appendix A: proof of theorem 1}
To distinguish those product states, some party has to start with a nontrivial orthogonality preserving measurement, say Alice. Since, the states are symmetrical, it is clear that if the states cannot be distinguished with Alice going first, then these states cannot be distinguished whether Bob or Charlie going first. Thus we only need to prove that these orthogonal product states cannot be distinguished by LOCC if Alice going first with orthogonality preserving positive-operator-valued  measurement.\\
Without loss of generality, suppose that Alice goes first with a set of $2d\times2d$  positive-operator-valued measurement(POVM) elements $\{{{M_{Ai}}^{\dagger}M_{Ai}}\}$. 
Let us consider $M_A$ be any matrix that preserves orthogonality of the above product states. Next we prove that ${M_{A}}^{\dagger}M_{A}\propto{\alpha I}$ for some $\alpha\in\mathbb{R}$.\\
We take states from FIG.2, by renaming $\ket{\psi_{i}}$ and $\ket{\phi_{i}^\pm}$, where $\ket{\psi_{i}}$ are given by,\\\\
$
\ket{\psi_{0}}=\ket{0}\ket{d}\ket{\overline{d-1}+d}\\  
\ket{\psi_{1}}=\ket{1}\ket{d}\ket{\overline{d-2}+\overline{d-1}}\\
\ket{\psi_{2}}=\ket{2}\ket{d}\ket{\overline{d-1}+d}\\ 
\ket{\psi_{3}}=\ket{3}\ket{d}\ket{\overline{d-2}+\overline{d-1}}\\
:\\
\ket{\psi_{d-1}}=\ket{d-1}\ket{d}\ket{\overline{d-1}+d}\\
\ket{\psi_{d}}=\ket{d}\ket{\overline{d-1}+d}\ket{d-1}\\
\ket{\psi_{d+1}}=\ket{d+1}\ket{d+\overline{d+1}}\ket{d-1}\\
\ket{\psi_{d+2}}=\ket{d+2}\ket{\overline{d-1}+d}\ket{d-1}\\
:\\:\\
\ket{\psi_{2d-1}}=\ket{2d-1}\ket{\overline{d-1}+d}\ket{d-1}$
\\\\
and the states $\ket{\phi_{i}^\pm}$ are given by,\\\\
$\ket{\phi_{0}^\pm}=\ket{0\pm1}\ket{0}\ket{d}\\
\ket{\phi_{1}^\pm}=\ket{1\pm2}\ket{1}\ket{d}\\
\ket{\phi_{2}^\pm}=\ket{2\pm3}\ket{2}\ket{d}\\
:\\:\\
\ket{\phi_{d-1}^\pm}=\ket{\overline{d-1}\pm d}\ket{d-1}\ket{d}\\
\ket{\phi_{d}^\pm}=\ket{d\pm\overline{d+1}}\ket{d-1}\ket{d+1}\\
:\\
:\\
:\\
\ket{\phi_{2d-2}^\pm}=\ket{\overline{2d-2}\pm\overline{2d-1}}\ket{d-1}\ket{2d-1}$\\\\
By using same methodology as we use previous, it can be shown that \\
$$0=\bra{\psi_{i}}{M_{A}}^{\dagger}M_{A}\otimes I \otimes I \ket{\psi_{j}}=m_{ij} ,\hspace{0.3 in}i\neq j$$\\ and
\begin{multline*} 
$$0=\bra{\phi_{i}^+}{M_{A}}^{\dagger}M_{A}\otimes I \otimes I \ket{\phi_{i}^-}=m_{ii}-m_{i,i+1}\\+m_{i+1,i}-m_{i+1,i+1},$$\\
$$0=\bra{\phi_{j}^+}{M_{A}}^{\dagger}M_{A}\otimes I \otimes I \ket{\phi_{j}^-}=m_{j,j}-m_{j,j+1}\\+m_{j+1,j}-m_{j+1,j+1},$$\\
$$1\leqslant i \leqslant d-1,d \leqslant j\leqslant 2d-2.
$$
\end{multline*}\\\\
Because of $m_{ij}=0$ for $i\neq j$, we get $m_{00=}=m_{11}=m_{22}=m_{33}=.....=m_{2d-1}$. Therefore, ${M_{A}}^{\dagger}M_{A}\propto{\alpha I}$ for some $\alpha\in\mathbb{R}$.\\
Since, $M_{A}$ is any arbitrary matrix that preserves orthogonality of the above product states. Thus for any measurement $\{{{M_{Ai}}^{\dagger}M_{Ai}}\}$ Alice can do only the trivial one. This implies the fact that the set of product states is LOCC indistinguishable.\hspace{1.7in} $\blacksquare$\\\\ 

\section*{Appendix B: proof of theorem 2}
Here, we prove the result by the method similar to the proof of Theorem 1. To distinguish those product states, some party has to start with a nontrivial orthogonality preserving measurement, say Alice. Since, the states are symmetrical, it is clear that if the states cannot be distinguished with Alice going first, then these states cannot be distinguished whether Bob or Charlie going first. Thus we only need to prove that these orthogonal product states cannot be distinguished by LOCC if Alice going first with orthogonality preserving positive-operator-valued  measurement.\\
Without loss of generality, suppose that Alice goes first with a set of $2d\times2d$  positive-operator-valued measurement(POVM) elements $\{{{M_{Ai}}^{\dagger}M_{Ai}}\}$. 
Let us consider $M_A$ be any matrix that preserves orthogonality of the above product states. Next we prove that ${M_{A}}^{\dagger}M_{A}\propto{\alpha I}$ for some $\alpha\in\mathbb{R}$.\\
We take states from FIG.2, by renaming $\ket{\psi_{i}}$ and $\ket{\phi_{i}^\pm}$, where $\ket{\psi_{i}}$ are given by,\\\\
$
\ket{\psi_{0}}=\ket{0}\ket{d}\ket{\overline{d-2}+\overline{d-1}}\\  
\ket{\psi_{1}}=\ket{1}\ket{d}\ket{\overline{d-1}+d}\\
\ket{\psi_{2}}=\ket{2}\ket{d}\ket{\overline{d-2}+\overline{d-1}}\\ 
\ket{\psi_{3}}=\ket{3}\ket{d}\ket{\overline{d-1}+d}\\
:\\
\ket{\psi_{d-1}}=\ket{d-1}\ket{d}\ket{\overline{d-1}+d}\\
\ket{\psi_{d}}=\ket{d}\ket{\overline{d-1}+d}\ket{d-1}\\
\ket{\psi_{d+1}}=\ket{d+1}\ket{d+\overline{d+1}}\ket{d-1}\\
\ket{\psi_{d+2}}=\ket{d+2}\ket{\overline{d-1}+d}\ket{d-1}\\
:\\:\\
\ket{\psi_{2d-1}}=\ket{2d-1}\ket{d+\overline{d+1}}\ket{d-1}$
\\\\
and the states $\ket{\phi_{i}^\pm}$ are given by,\\\\
$\ket{\phi_{0}^\pm}=\ket{0\pm1}\ket{0}\ket{d}\\
\ket{\phi_{1}^\pm}=\ket{1\pm2}\ket{1}\ket{d}\\
\ket{\phi_{2}^\pm}=\ket{2\pm3}\ket{2}\ket{d}\\
:\\
\ket{\phi_{d-1}^\pm}=\ket{\overline{d-1}\pm d}\ket{d-1}\ket{d}\\
\ket{\phi_{d}^\pm}=\ket{d\pm\overline{d+1}}\ket{d-1}\ket{d+1}\\
:\\
:\\
\ket{\phi_{2d-2}^\pm}=\ket{\overline{2d-2}\pm\overline{2d-1}}\ket{d-1}\ket{2d-1}\\
$\\\\
By using same methodology as we use previously, it can be shown that \\
$$0=\bra{\psi_{i}}{M_{A}}^{\dagger}M_{A}\otimes I \otimes I \ket{\psi_{j}}=m_{ij} ,\hspace{0.3 in}i\neq j$$\\ and
\begin{multline*} 
$$0=\bra{\phi_{i}^+}{M_{A}}^{\dagger}M_{A}\otimes I \otimes I \ket{\phi_{i}^-}=m_{ii}-m_{i,i+1}\\+m+_{i+1,i}-m_{i+1,i+1},$$\\
$$0=\bra{\phi_{j}^+}{M_{A}}^{\dagger}M_{A}\otimes I \otimes I \ket{\phi_{j}^-}=m_{j,j}-m_{j,j+1}\\+m_{j+1,j}-m_{j+1,j+1},$$\\
$$1\leqslant i \leqslant d-1,d \leqslant j\leqslant 2d-2.
\end{multline*}\\
Because of $m_{ij}=0$ for $i\neq j$, we get $m_{00=}=m_{11}=m_{22}=m_{33}=.....=m_{2d-1}$. Therefore, ${M_{A}}^{\dagger}M_{A}\propto{\alpha I}$ for some $\alpha\in\mathbb{R}$.\\
Since, $M_{A}$ is any arbitrary matrix that preserves orthogonality of the above product states. Thus for any measurement $\{{{M_{Ai}}^{\dagger}M_{Ai}}\}$ Alice can do only the trivial one. This implies the fact that the set of product states is LOCC indistinguishable.\hspace{1.7in} $\blacksquare$\\
\section*{Appendix C: proof of Example 2}
For distinguishing the states, again, some party has to start with a nontrivial and non-disturbing measurement, i.e., not all measurements $\{{{M_{Ai}}^{\dagger}M_{Ai}}\}$ are proportional to the identity and have the orthogonality relations preserved afterwards, making further discrimination possible.\\ 
Suppose, Alice goes first with a set of $5\times5$  positive-operator-valued measurement(POVM) elements $\{{{M_{Ai}}^{\dagger}M_{Ai}}\}$. 
Let us consider $M_{A}$ be any matrix that preserves orthogonality of the above product states. Where $M_A$ has the following representation under the basis $\{\ket{0},\ket{1},..,\ket{4}\}$\\
$${M_{A}}^{\dagger}M_{A}=\begin{bmatrix}
m_{00} & m_{01} & . & . & m_{04}\\
m_{10} & m_{11} & . & . & m_{14}\\
. & . & .  &  & .\\
. & . &   & . & .\\
m_{40} & m_{51} &.&.& m_{44}
\end{bmatrix}$$\\
Next, we prove that ${M_A}^{\dagger}M_A\propto{\alpha I}$.\\\\
After performing Alice's measurement, the post-measurement states $\left\lbrace{M_A\otimes I_B\otimes I_C\ket{\phi_i}, i = 1, . . . ,31}\right\rbrace $
should be mutually orthogonal. Considering the
states $\ket{\phi_{5,21,25,26,27}}$, we know that $\bra{\phi_{21}}{M_{A}}^{\dagger}M_{A}\otimes I_B \otimes I_C \ket{\phi_{25}}=0$. Then $\bra{0}{M_{A}}^{\dagger}M_{A}\ket{1} \bra{2}\ket{2}\bra{2}\ket{2+3}=0$. So, $\bra{0}{M_{A}}^{\dagger}M_{A}\ket{1}=0$, i.e., $m_{01}=m_{10}=0$. In the same way, by using the other states we get $m_{ij}=0, i,j=0,1,2,3,4, i\neq j$.\\	For the states  $\ket{\phi_{15,16}}$ we have, $\bra{0+1}{M_{A}}^{\dagger}M_{A}\ket{0-1} \bra{0}\ket{0}\bra{2}\ket{2}=0$, i.e., $\bra{0}{M_{A}}^{\dagger}M_{A}\ket{0}-\bra{1}{M_{A}}^{\dagger}M_{A}\ket{1}=0$. Thus, $m_{00}=m_{11}$. Using the same fact for the states $\ket{\phi_{13,14}}$, $\ket{\phi_{1,2}}$, $\ket{\phi_{3,4}}$ we get $m_{22}=m_{33}$, $m_{11}=m_{22}$, $m_{33}=m_{44}$ respectively. Thus finally, we get $m_{00}=m_{11}=m_{22}=m_{33}=m_{44}$.\\
Therefore, all of Alice's measurements ${M_A}^{\dagger}M_A$ are proportional to the identity, meaning that Alice cannot start with a nontrivial measurement. As the set of states has a symmetrical representation then it can also be showed that all measurements of Bob and Charlie should be proportional to identity. Thus, no one can start with a nontrivial measurement. Hence, the set of states cannot be perfectly distinguished by LOCC only.\hspace{0.1in} $\blacksquare$\\
\section*{Appendix D: Bell state discrimination}
Example 4: In      ${\mathbb{C}}^{6}\bigotimes{\mathbb{C}}^{6}\bigotimes{\mathbb{C}}^{6}$ the set of 36 orthogonal product states\\
$$\begin{Bmatrix}
	\begin{matrix}
		\ket{\psi_{1}}=\ket{2+3}\ket{0}\ket{3}\\
		\ket{\psi_{2}}=\ket{3}\ket{2+3}\ket{0}\\
		\ket{\psi_{3}}=\ket{0}\ket{3}\ket{2+3}
	\end{matrix} & 
	\begin{matrix}
		\ket{\psi_{4}}=\ket{2+3}\ket{2}\ket{5}\\
		\ket{\psi_{5}}=\ket{5}\ket{2+3}\ket{2}\\
		\ket{\psi_{6}}=\ket{2}\ket{5}\ket{2+3}
	\end{matrix}\\\\
	\begin{matrix}
		\ket{\psi_{7,8}}=\ket{2\pm3}\ket{2}\ket{3}\\
		\ket{\psi_{9,10}}=\ket{3}\ket{2\pm3}\ket{2}\\
		\ket{\psi_{11,12}}=\ket{2}\ket{3}\ket{2\pm3}
	\end{matrix} & 
	\begin{matrix}
		\ket{\psi_{13,14}}=\ket{3\pm4}\ket{2}\ket{4}\\
		\ket{\psi_{15,16}}=\ket{4}\ket{3\pm4}\ket{2}\\
		\ket{\psi_{17,18}}=\ket{2}\ket{4}\ket{3\pm4}
	\end{matrix}\\\\ 
	\begin{matrix} 
		\ket{\psi_{19,20}}=\ket{1\pm2}\ket{1}\ket{3}\\
		\ket{\psi_{21,22}}=\ket{3}\ket{1\pm2}\ket{1}\\
		\ket{\psi_{23,24}}=\ket{1}\ket{3}\ket{1\pm2}
	\end{matrix} & 
	\begin{matrix}
		\ket{\psi_{25,26}}=\ket{0\pm1}\ket{0}\ket{3}\\
		\ket{\psi_{27,28}}=\ket{3}\ket{0\pm1}\ket{0}\\
		\ket{\psi_{29,30}}=\ket{0}\ket{3}\ket{0\pm1}
	\end{matrix}\\\\
	\begin{matrix} 
		\ket{\psi_{31,32}}=\ket{4\pm5}\ket{2}\ket{5}\\
		\ket{\psi_{33,34}}=\ket{5}\ket{4\pm5}\ket{2}\\
		\ket{\psi_{35,36}}=\ket{2}\ket{5}\ket{4\pm5}
	\end{matrix}\\\\
\end{Bmatrix}$$\\ can be distinguished by LOCC with one-copy of Bell state shared between any two parties. But there is a restriction on that protocol, one of them who share the Bell state have to start.\\\\\\
Proof: First of all, let any two parties say, Alice and Bob share a Bell state $\ket{\psi}_{ab}=\frac{1}{\sqrt{2}}(\ket{00}+\ket{11})$. Alice performs a two-outcome measurement where each outcome corresponds to a rank-6 projector:
\begin{multline*}
	A_1=\ket{00}_{Aa}\bra{00}+\ket{10}_{Aa}\bra{10}+\ket{20}_{Aa}\bra{20}\\
	+\ket{31}_{Aa}\bra{31}+\ket{41}_{Aa}\bra{41}+\ket{51}_{Aa}\bra{51}
\end{multline*}
\begin{multline}
	$$A_2=\ket{01}_{Aa}\bra{01}+\ket{11}_{Aa}\bra{11}+\ket{21}_{Aa}\bra{21}\\
	+\ket{30}_{Aa}\bra{30}+\ket{40}_{Aa}\bra{40}+\ket{50}_{Aa}\bra{50}$$
\end{multline}
After operating the projector $C_1$ on systems $Cc$, each of the initial states
is transformed into\\
$$\begin{pmatrix}
	\begin{Bmatrix}
		\begin{matrix}
			\ket{0}\ket{3}\ket{0\pm1}\\
			\ket{0\pm1}\ket{0}\ket{3}\\ 
			\ket{1}\ket{3}\ket{1\pm2}\\
			\ket{1\pm2}\ket{1}\ket{3}\\
			\ket{2}\ket{4}\ket{3\pm4}
		\end{matrix} & 
		\begin{matrix}
			\ket{2}\ket{3}\ket{2\pm3}\\
			\ket{2}\ket{5}\ket{4\pm5}\\
			\ket{2}\ket{5}\ket{2+3}\\
			\ket{0}\ket{3}\ket{2+3}
		\end{matrix}
	\end{Bmatrix}\bigotimes\ket{00}_{ab}$$\\
	$$\begin{Bmatrix}
		\begin{matrix}
			\ket{3}\ket{0\pm1}\ket{0}\\
			\ket{3}\ket{2+3}\ket{0}\\
			\ket{5}\ket{4\pm5}\ket{2}\\ 
			\ket{3}\ket{1\pm2}\ket{1}\\
			\ket{5}\ket{2+3}\ket{2}
		\end{matrix} & 
		\begin{matrix}
			\ket{3\pm4}\ket{2}\ket{4}\\
			\ket{3}\ket{2\pm3}\ket{2}\\
			\ket{4}\ket{3\pm4}\ket{2}\\
			\ket{4\pm5}\ket{2}\ket{5}
		\end{matrix}
	\end{Bmatrix}\bigotimes\ket{11}_{ab}$$\\
	
	$$\begin{Bmatrix}
		\ket{203}_{ABC}\bigotimes\ket{00}_{ab}+\ket{303}_{ABC}\bigotimes\ket{11}_{ab}\\\ket{225}_{ABC}\bigotimes\ket{00}_{ab}+\ket{325}_{ABC}\bigotimes\ket{11}_{ab}\\
		\ket{223}_{ABC}\bigotimes\ket{00}_{ab}\pm\ket{323}_{ABC}\bigotimes\ket{11}_{ab}
	\end{Bmatrix}
\end{pmatrix}$$\\
i.e.,
\begin{multline}
	\ket{{\psi\prime}_{i}}=\ket{\psi_{i}}_{ABC}\bigotimes\ket{00}_{ab},  i=3,6,11,12,17,18,\\
	19,20,23,24,25,26,29,30,35,36\\
	\ket{{\psi\prime}_{i}}=\ket{\psi_{i}}_{ABC}\bigotimes\ket{11}_{ab},  i=2,5,9,10,13,14,\\
	15,16,21,22,27,28,31,32,33,34.\\
	\ket{{\psi\prime}_{3}}=\ket{203}_{ABC}\bigotimes\ket{00}_{ab}+\ket{303}_{ABC}\bigotimes\ket{11}_{ab}\\
	\ket{{\psi\prime}_{6}}=\ket{225}_{ABC}\bigotimes\ket{00}_{ab}+\ket{325}_{ABC}\bigotimes\ket{11}_{ab}\\
	\ket{{\psi\prime}_{11,12}}=\ket{223}_{ABC}\bigotimes\ket{00}_{ab}\pm\ket{323}_{ABC}\bigotimes\ket{11}_{ab}
\end{multline}\\
After operating the projector $A_2$ on systems $Aa$, it creates new states
which differ from the states $(5)$ only by ancillary systems
$\ket{00}_{ab}\rightarrow\ket{11}_{ab}$ and $\ket{11}_{ab}\rightarrow\ket{00}_{ab}$. Then the latter can be handled using the exact same method as for $A_1$. Thus, we only
need to discuss the behaviour after operating $A_1$.
Now we describe the method to distinguish those states how the parties act. After $A_1$ clicks, Bob makes  five-outcome projective measurement. The first outcome $B_1=\ket{1}_B\bra{1}\bigotimes\ket{0}_b\bra{0}$, which leaves $\ket{{\psi\prime}_{19,20}}$ and remain same for all other states in $(2)$. Then Alice can easily discriminate the two states by projecting onto $\ket{1\pm2}_A$. The second outcome is $B_2=\ket{4}_B\bra{4}\bigotimes\ket{0}_b\bra{0}$. The only remaining possibility is $\ket{{\psi\prime}_{17,18}}$, which can be successfully identified by Alice by projecting onto  $\ket{3\pm4}_A$. Using rank one projector $B_3=\ket{3}_B\bra{3}\bigotimes\ket{0}_b\bra{0}$ on Bob's Hilbert space leaves $\ket{{\psi\prime}_{3,11,12,23,24,29,30}}$ and annihilates other states. Corresponding to $B_3$ when Alice uses projector $A_{31}=\ket{0}_A\bra{0}\bigotimes\ket{0}_a\bra{0}$, it leaves $\ket{{\psi\prime}_{3,29,30}}$, which can be easily distinguished by projectors  $\ket{0\pm1}_C$ and $\ket{2+3}_C$. When Alice uses projector $A_{32}=\ket{1}_A\bra{1}\bigotimes\ket{0}_a\bra{0}$, it leaves $\ket{{\psi\prime}_{23,24}}$, which can be easily distinguished by projectors  $\ket{1\pm2}_C$. Also when Alice uses projector $A_{33}=\ket{2}_A\bra{2}\bigotimes\ket{0}_a\bra{0}$, it leaves $\ket{{\psi\prime}_{11,12}}$, which can be easily distinguished by projectors  $\ket{2\pm3}_C$. The fourth outcome is $B_4=\ket{5}_B\bra{5}\bigotimes\ket{0}_b\bra{0}$. The only remaining possibility is $\ket{{\psi\prime}_{6,35,36}}$, which can be successfully identified by Charlie by projecting onto  $\ket{2+3}_C$ and $\ket{4\pm5}_C$.\\\\
Bob's last outcome is $B_5=I_B-\left( B_1+B_2+B_3+B_4\right) $, which leaves all other states excepting $14$ states which has been discriminated above. Now corresponding to $B_5$, Charlie makes $6$-outcome projective measurement. The first outcome $C_{51}=\ket{0}_C\bra{0}$  which leaves $\ket{{\psi\prime}_{2,27,28}}$ and remain same for all other states. If Bob projects onto $\ket{0\pm1}$ and $\ket{2+3}$, then all states are successfully identified. Now, if $C_{52}=\ket{1}_C\bra{1}$ occur, it leaves $\ket{{\psi\prime}_{21,22}}$ and thus successfully identified by Bob by projecting onto $\ket{1\pm2}$. Charlie's third outcome $C_{53}=\ket{2}_C\bra{2}$  which leaves $\ket{{\psi\prime}_{5,9,10,15,16,33,34}}$ and remain same for all other states. Again, if Alice projects onto $\ket{3}_A\bra{3}\bigotimes\ket{1}_a\bra{1}$, it leaves two states $\ket{{\psi\prime}_{9,10}}$ and thus successfully distinguished. If Alice projects onto $\ket{4}_A\bra{4}\bigotimes\ket{1}_a\bra{1}$ which leaves $\ket{{\psi\prime}_{15,16}}$ and thus Bob will distinguish by projecting onto $\ket{3\pm4}_B$. Also, if Alice projects onto 
$\ket{5}_A\bra{5}\bigotimes\ket{1}_a\bra{1}$ which leaves $\ket{{\psi\prime}_{5,33,34}}$ and thus Bob distinguish $\ket{{\psi\prime}_{33,34}}$ by projecting onto $\ket{4\pm5}_B$ and $\ket{{\psi\prime}_{5}}$ by projecting onto $\ket{2+3}_B$. Charlie's fourth outcome $C_{54}=\ket{3}_C\bra{3}$  which leaves $\ket{{\psi\prime}_{1,7,8,25,26}}$ and remain same for all other states. Again, if Alice projects onto $\ket{0\pm1}_A\bra{0\pm1}\bigotimes\ket{0}_a\bra{0}$, it successfully distinguished two states $\ket{{\psi\prime}_{25,26}}$. If Alice projects onto $\ket{2}_A\bra{2}\bigotimes\ket{0}_a\bra{0}+\ket{3}_A\bra{3}\bigotimes\ket{1}_a\bra{1}$ which leaves $\ket{{\psi\prime}_{1,7,8}}$ and thus Bob will distinguish $\ket{{\psi\prime}_{1}}$ by projecting onto $\ket{0}_B\bra{0}\bigotimes\ket{0}_b\bra{0}+\ket{0}_B\bra{0}\bigotimes\ket{1}_b\bra{1}$ and distinguish $\ket{{\psi\prime}_{7,8}}$ by projecting onto $\ket{2}_B\bra{2}\bigotimes\ket{0}_b\bra{0}+\ket{2}_B\bra{2}\bigotimes\ket{1}_b\bra{1}$.Charlie's fifth outcome $C_{55}=\ket{4}_C\bra{4}$  which leaves $\ket{{\psi\prime}_{13,14}}$ and remain same for all other states, which can be easily distinguished by Alice by projecting onto $\ket{3\pm4}_A$.\\\\
Charlie's last outcome is $C_{56}=I_C-\left( C_{51}+C_{52}+C_{53}+C_{54}+C_{55}\right) $, which leaves $\ket{{\psi\prime}_{4,31,32}}$. Thus Alice will distinguish $\ket{{\psi\prime}_{4}}$ by projecting onto $\ket{2}_A\bra{2}\bigotimes\ket{0}_a\bra{0}+\ket{3}_A\bra{3}\bigotimes\ket{1}_a\bra{1}$ and distinguish $\ket{{\psi\prime}_{31,32}}$ by projecting onto $\ket{4\pm5}_A\bra{4\pm5}\bigotimes\ket{1}_a\bra{1}$.  This completes our argument and proof of the result.\hspace{2.5in}$\blacksquare$\\
As similar approch it can be shown that those class of states contains $18(d-1)$ states described in Theorem 1 and Theorem 2 can be perfectly distinguished by LOCC if any two parties share a Bell state. But there is a restriction on that protocol, one of them who share the Bell state have to start the protocol. But, if a GHZ state is used as a resource then any one can start the discrimination protocol.\\

Theorem 9: In  ${\mathbb{C}}^{2k+1}\bigotimes{\mathbb{C}}^{2l+1}\bigotimes{\mathbb{C}}^{2m+1}$, 2-copy of Bell states state shared between any two parties is sufficient to perfectly distinguished the set of $6(k+l+m)-5$ orthogonal product states by LOCC.\\
\textit{Proof:} First of all, let Alice and Bob share a Bell state $\ket{\psi}_{ab}=\frac{1}{\sqrt{2}}(\ket{00}+\ket{11})$. Then Alice will perform a two-
outcome measurement, each outcome corresponding to a rank-(2k-1) projector:\\
\begin{multline*}
	$$A_1=\ket{00}_{Aa}\bra{00}+\ket{10}_{Aa}\bra      {10}+\ket{20}_{Aa}\bra{20}+.......\\+\ket{(2m-1)0}_{Aa}\bra{(2m-1)0}
	+\ket{(2m)1}_{Aa}\bra{(2m)1}
\end{multline*}
\begin{multline*}
	A_2=\ket{01}_{Aa}\bra{01}+\ket{11}_{Aa}\bra{11}+\ket{21}_{Aa}\bra{21}+......\\
	+\ket{(2m-1)1}_{Aa}\bra{(2m-1)1}
	+\ket{(2m)0}_{Aa}\bra{(2m)0}$$
\end{multline*}
After operating the projector $A_1$ on systems $Aa$, each of the initial states
will transform into\\ 
\begin{widetext}
	\begin{multline*}
		\ket{{\phi\prime}_{i}}=\ket{\phi_{i}}_{ABC}\bigotimes\ket{00}_{ab} ,  i=1,2,...,(2k-2),(2k+2m+1),...,(6k+6l+6m-5)\\
		\ket{{\phi\prime}_{i}}=\ket{\phi_{i}}_{ABC}\bigotimes\ket{11}_{ab} ,  i=(2k+1),(2k+2),...(2k+2m)\\
		\ket{{\phi\prime}_{2k-1,2k}}=\ket{(2m-1)(2k)(l)}_{ABC}\bigotimes\ket{00}_{ab}+\ket{(2m)(2k)(l)}_{ABC}\bigotimes\ket{11}_{ab}\\
	\end{multline*}\\
\end{widetext}
If $A_1$ click, Bob makes  two-outcome projective measurement. The first outcome $B_1=\ket{l}_B\bra{l}\bigotimes\ket{1}_b\bra{1}$, which leaves $\ket{{\phi\prime}_{(2k+1),(2k+2),...(2k+2m)}}$ and those can be distinguished by projecting onto $Cc$ system. If $B_2=I-B_1$ click, it isolates $\ket{{\phi\prime}_{1,2,..,2k,2k+2m+1,2k+2m+2,..,6(k+l+m)-5}}$. After $B_2$ click, Charlie makes  three-outcome projective measurement, $C_1=\ket{2m}_C\bra{2m}$, $C_2=\ket{0}_C\bra{0}+\ket{1}_C\bra{1}$, $C_3=I-C_1-C_2$. If $C_1$ clicks, it isolates $\ket{{\phi\prime}_{2k+2m+1,2k+2m+2,...2k+2l+2m}}$, which can be easily distinguished by projecting onto the $Bb$ system. If  $C_2$ clicks, it isolates $\ket{{\phi\prime}_{4k+2l+2m+1,..,4k+4l+2m-2,4k+4l+4m-1,4k+4l+4m,6k+6l+4m-2}}$, which can be easily distinguished by Alice and Bob. If  $C_3$ clicks, it isolates the remaining all other states. Now to distinguish the remaining states corresponding to the outcome $C_3$, a Bell state shared between any two parties is sufficient. The procedure is same as we have described above. That is, we have succeeded in designing a protocol to distinguish the states by LOCC with 2-copy of Bell states. This completes the proof.\hspace{2.0in}$\blacksquare$

Similarly, by the protocol we have described in the previous theorem, we can show that 2-copy of Bell states shared between any two parties is sufficient to distinguish each of the above classes in ${\mathbb{C}}^{2k+1}\bigotimes{\mathbb{C}}^{2l+1}\bigotimes{\mathbb{C}}^{2m}$, ${\mathbb{C}}^{2k+1}\bigotimes{\mathbb{C}}^{2l}\bigotimes{\mathbb{C}}^{2m}$,\\
${\mathbb{C}}^{2k}\bigotimes{\mathbb{C}}^{2l}\bigotimes{\mathbb{C}}^{2m}$ (Th.3 - Th.6).\\\\


\begin{thebibliography} {100}
\bibitem{1} Bennett, C.H., DiVincenzo, D.P., Fuchs, C.A., Mor, T., Rains, E., Shor, P.W., Smolin, J.A., Wootters, W.K.: Phys. Rev. A. 59, 1070 (1999) 
\bibitem{2} Bennett, C.H., DiVincenzo, D.P., Mor, T., Rains, E., Shor, P.W., Smolin, J.A., Terhal, B.M.: Phys. Rev. Lett. 82, 5385 (1999)
\bibitem{3} DiVincenzo, D.P., Mor, T., Shor, P.W., Smolin, J.A., Terhal, B.M.: Communications in Mathematical Physics 238, 379 (2003)
\bibitem{4} Walgate, J., Hardy, L.: Phys. Rev. Lett. 89, 147901 (2002)
\bibitem{5} Niset, J., Cerf, N.J.: Phys. Rev. A. 74, 052103 (2006)
\bibitem{6} Zhang, Z.-C., Gao, F., Qin, S.-J., Yang, Y.-H., Wen, Q.-Y.: Phys. Rev. A. 90, 022313 (2014)
\bibitem{7} Bandyopadhyay, S., Walgate, J.: J. Phys. A: Math. Theor. 42, 072002 (2009)
\bibitem{8} Wang, Y.-L., Li, M.-S., Zheng, Z.-J., Fei, S.-M.: Phys. Rev. A 92, 032313 (2015)
\bibitem{9} Chen, J., Johnston, N.: Communications in Mathematical Physics 333, 351 (2015)
\bibitem{10} Zhang, Z.-C., Gao, F., Cao, Y., Qin, S.-J., Wen, Q.-Y.: Phys. Rev. A 93, 012314 (2016)
\bibitem{11} Xu, G.-B., Wen, Q.-Y., Qin, S.-J., Yang, Y.-H., Gao, F.: Phys. Rev. A 93, 032341 (2016)
\bibitem{12} Zhang, X., Tan, X., Weng, J., Li, Y.: Sci. Rep. 6, 28864 (2016)
\bibitem{13} Wang, Y.-L., Li, M.-S., Fei, S.-M., Zheng, Z.-J.: arXiv:1703.06542 [quant-ph] (2017)
\bibitem{14} Zhang, Z.-C., Zhang, K.-J., Gao, F., Wen, Q.-Y., Oh, C.H.: Phys. Rev. A 95, 052344 (2017)
\bibitem{15} Wang, Y.-L., Li, M.-S., Zheng, Z.-J., Fei, S.-M.: Quantum Information Processing 16, 5 (2016)
\bibitem{16} Zhang, X., Weng, J., Tan, X., Luo, W.: Quantum Information Processing 16, 168 (2017)
\bibitem{17} Halder, S.: Phys. Rev. A 98, 022303 (2018)
\bibitem{18} Halder, S., Banik, M., Ghosh, S.: arXiv:1801.00405 [quant-ph] (2018)
\bibitem{19} Halder, S., Banik, M., Agrawal, S., Bandyopadhyay, S.: Phys. Rev. Lett. 122, 040403 (2019)
\bibitem{20} Rout, S., Maity, A.G., Mukherjee, A., Halder, S., Banik, M.: arXiv:1905.05930 [quant-ph] (2019)
\bibitem{21} Cohen, S.M.: Phys. Rev. A 75, 052313 (2007)
\bibitem{22} Cohen, S.M.: Phys. Rev. A 77, 012304 (2008)
\bibitem{23}  Bandyopadhyay, S., Rahaman, R., Wootters, W.K.: Journal of Physics A: Mathematical and Theoretical 43, 455303 (2010)
\bibitem{24} Yu, N., Duan, R., Ying, M.: IEEE Transactions on Information Theory 60, 2069 (2014)
\bibitem{25} Bandyopadhyay, S., Cosentino, A., Johnston, N., Russo, V., Watrous, J., Yu, N.: IEEE Transactions on Information Theory 61, 3593 (2015)
\bibitem{26} Bandyopadhyay, S., Halder, S., Nathanson, M.: Phys. Rev. A 94, 022311 (2016)
\bibitem{27} Zhang, Z.-C., Gao, F., Cao, T.-Q., Qin, S.-J., Wen, Q.-Y.: Sci. Rep. 6, 30493 (2016)
\bibitem{28} Bandyopadhyay, S., Halder, S., Nathanson, M.: Phys. Rev. A 97, 022314 (2018)
\bibitem{29} Zhang, Z.-C., Song, Y.-Q., Song, T.-T., Gao, F., Qin, S.-J., Wen, Q.-Y.: Phys. Rev. A 97, 022334 (2018)
\bibitem{30} Li, L.-J., Gao, F., Zhang, Z.-C., Wen, Q.-Y.: Phys. Rev. A 99, 012343 (2019)
\bibitem{31} Bennett, C.H., Brassard, G., Crépeau, C., Jozsa, R., Peres, A., Wootters, W. K.: Phys. Rev. Lett. 70, 1895 (1993)
\bibitem{32} Xu, G.-B., Wen, Q.-Y., Qin, S.-J., Yang, Y.-H., Gao, F.: Phys. Rev. A 93, 032341 (2016)
\bibitem{33} Wang, Y.-L., Li, M.-S., Zheng, Z.-J., Fei, S.-M.: Phys. Rev. A 92, 032313 (2015)
\bibitem{34} Sengupta, R., Arvind: Phys. Rev. A 87, 012318 (2013)
\bibitem{35} Halder, S., Sengupta, R.: Physics Letters A 383, 2004 (2019)
\bibitem{36}  Walgate, J., Short, A.J.,  Hardy, L., Vedral, V.: Phys. Rev. Lett. 85, 4972 (2000)
\bibitem{37}  Virmani, S., Sacchi, M.F., Plenio, M.B., Markham, D.: Phys. Lett. A. 288, 62 (2001)
\bibitem{38} Ghosh, S., Kar, G., Roy, A., Sen(De), A., Sen, U.: Phys. Rev. Lett. 87, 277902 (2001)
\bibitem{39} Horodecki, M., Sen(De), A., Sen, U., Horodecki, K.: Phys. Rev. Lett 90, 047902 (2003)
\bibitem{40} Ghosh, S., Kar, G., Roy, A., Sarkar, D.: Phys. Rev. A. 70, 022304 (2004)
\bibitem{41} Nathanson, M.: J. Math. Phys. 46, 062103 (2005)
\bibitem{42} Watrous, J.: Phys. Rev. Lett. 95, 080505 (2005)
\bibitem{43} Fan, H.: Phys. Rev. A 75, 014305 (2007)
\bibitem{44} Yu, N., Duan, R., Ying, M.: Phys. Rev. Lett. 109, 020506 (2012)
 	 

 
\end{thebibliography}
\end{document}